\documentclass[lettersize,journal]{IEEEtran}
\usepackage{amsmath,amsfonts,amssymb,amsthm}
\usepackage{multirow}
\usepackage{epstopdf}
\usepackage[linesnumbered,lined,boxed,commentsnumbered,ruled]{algorithm2e} 
\usepackage{setspace}
\usepackage{threeparttable}
\usepackage{array}
\usepackage[caption=false,font=footnotesize,labelfont=rm,textfont=rm]{subfig}
\usepackage{textcomp}
\usepackage{stfloats}
\usepackage{url}
\usepackage{verbatim}
\usepackage{booktabs}
\usepackage[pdftex]{graphicx}
\usepackage{epstopdf}
\usepackage{array,tabularx}
\usepackage{tikz}
\usepackage{cite}
\usepackage{float}
\usepackage{bm}
\begin{document}
	\title{Analysis of Hierarchical AoII over Unreliable Channels: A Stochastic Hybrid System Approach}
	%\author{IEEE Publication Technology Department
		%\thanks{Manuscript created October, 2020; This work was developed by the IEEE Publication Technology Department. This work is distributed under the \LaTeX \ Project Public License (LPPL) ( http://www.latex-project.org/ ) version 1.3. A copy of the LPPL, version 1.3, is included in the base \LaTeX \ documentation of all distributions of \LaTeX \ released 2003/12/01 or later. The opinions expressed here are entirely that of the author. No warranty is expressed or implied. User assumes all risk.}}
	%
	%\markboth{Journal of \LaTeX\ Class Files,~Vol.~18, No.~9, September~2020}%
	%{How to Use the IEEEtran \LaTeX \ Templates}
	\author{Han~Xu, Jiaqi~Li, Jixiang~Zhang,
		Tiecheng~Song,
		and~Yinfei~Xu,
		
		\thanks{This work was supported in part by the National Natural Science Foundation of China under Grant 62371119, and in part by the Open Research Fund of the State Key Laboratory of Integrated Services Networks, Xidian University. The authors are with the National Mobile Communications Research Laboratory, Southeast University, Nanjing 210096, China. Han Xu and Yinfei Xu are also with the State Key Laboratory of Integrated Services Networks, Xidian University,China  (e-mail: \{han\_xu, jiaqili, zhangjx, songtc, yinfeixu\}@seu.edu.cn).} % <-this % stops a space
		%	\thanks{Hao~Xu is with the Department of Electronic and Electrical Engineering, University College London, London WC1E7JE, UK (e-mail: hao.xu@ucl.ac.uk).}
	}%
	\maketitle
	
	\begin{abstract}
		In this work, we generalize the Stochastic Hybrid Systems (SHSs) analysis of Age of information (AoI) to the Age of Incorrect Information (AoII) metric. Hierarchical ageing processes are adopted using the continuous AoII for the first time. Two different hierarchy schemes are considered: 1). A hierarchy of zero and linear ageing processes with different slopes; 2). A hierarchy of zero, linear and exponential ageing processes. We first modify the main result in \cite[Theorem 1]{yates_age_2020b} to provide a systematic way to analyze the continuous hierarchical AoII over unslotted real-time systems. The closed-form expressions of average hierarchical AoII are obtained in two typical scenarios with different channel conditions, i.e., an M/M/1/1 queue over noisy channels and two M/M/1/1 queues over collision channels. Moreover,  under each scenario, we analyze the stability issue regarding positive recurrence and provide the stability conditions that ensure a steady-state  average AoII. Finally, we compare the closed-form results between average AoI and  AoII in the M/M/1/1 queue. The effects of different channel parameters on the average hierarchical AoII are also evaluated. 
	\end{abstract}
	

	\section{Introduction}
	\IEEEPARstart{I}{n} real-time status update systems, the study of information freshness over the network layer has been the subject of ongoing interest in the past decade. A classic measure, Age of Information (AoI), was first introduced to measure information freshness. The AoI $\Delta(t)$ measured by the monitor at time $t$ is defined by $\Delta(t)=t-U(t)$, which is the elapsed time since the generation time $U(t)$ of the last received update.     Since then, extensively researches have focused on two dominant directions in the AoI field: i). The analysis of closed-form average AoI (and its moments) over queueing and other networking systems, e.g., \cite{costa_age_2016, kam_effect_2016, kam_age_2018, inoue_general_2019, yates_age_2019, yates_age_2020b, najm_content_2020, buyukates_version_2022}; ii).  The optimization of AoI performance that searches for the optimal scheduling policies over different networking systems, e.g., \cite{sun_update_2017, i_kadota_scheduling_2018-1, bedewy_age_2019, hsu_scheduling_2020, maatouk_age_2020-1, chen_age_2022,yao_age-optimal_2022}. A comprehensive survey of these works can be found in \cite{sun_age}.
	
	However, the major drawback of AoI is that this metric only considers the time aspect of update but is content-agnostic. The AoI process keeps growing unless a new update is received. However, such an ageing process may not suitable for all the real-time update systems. For example, if the newly-generated source update contains the same information as the one stored at the monitor, the ageing process of the system should be  ``paused'' for a while since the current update is still fresh enough to be used.  Therefore, a variety of modified metrics have been proposed to address this content-agnostic issue. To our best knowledge, these new metrics include Effective AoI (EAoI) \cite{kam_towards_2018}, Age of Synchronization (AoS)\cite{zhong_two_2018}, Query AoI (QAoI)\cite{chiariotti_query_2022}, binary freshness\cite{bastopcu_gossiping_2021}, Urgency of Information (UoI)\cite{zheng_urgency_2020}, Age of Incorrect Information (AoII) \cite{maatouk_age_2020}, etc.\footnote{There are also several variations of AoII, e.g., Age of Changed Information (AoCI)\cite{lin_average_2020},  Age of Incorrect Estimates(AoIE)\cite{joshi_minimization_2021}, Age of Processed Information (AoPI)\cite{jayanth_age_2023}.} These metrics not only include the age perspective but also consider the content-mismatch between the updates at the source and monitor. In this work, we adopt the most representative form among them, i.e., AoII, and extend the SHSs analysis of AoI \cite{yates_age_2019} into the AoII metric. First, we give the definition of AoII first proposed in \cite{maatouk_age_2020}, i.e.,
	\begin{equation}\label{gAoII}
		x(t) =f(t)\times g(X(t), \hat{X}(t)),
	\end{equation} 
	where function $f:[0, \infty)\to[0, \infty)$ is an increasing age penalty function paid for being unaware of the correct status of the source process. Function $g:\Omega\times\Omega\to[0, \infty)$ is an information penalty function that reflects the content-mismatch between the current estimate $\hat{X}(t)$ at the monitor and the actual content $X(t)$ of the process. Hence, AoII is identified as a content-aware age metric and also a promising alternative measurement of information freshness  compared to AoI. There has been a growing interest in studying the average AoII performance and optimal scheduling policy, e.g., \cite{kam_age_2020, kriouile_minimizing_2021, maatouk_age_2023, nayak_decentralized_2023, ayik_optimization_2023, meng_toward_2023, chen_minimizing_2024, cosandal_aoii-optimum_2024, bountrogiannis_age_2025}.  However,  our work differs from them in two essential ways:
	
	1) We extend the AoII metric into unslotted real-time update systems using a continuous form of \eqref{gAoII}, which, to our best of knowledge, has only been used in  \cite{cosandal_aoii-optimum_2024}. Furthermore, we  pursue the closed-form expressions of AoII instead of an optimal policy by first introducing Stochastic Hybrid Systems (SHSs) into the analysis of content-aware AoII process. 
	
	2) Since the level of age dissatisfaction measured  in \eqref{gAoII} is determined by different choices for functions $f$ and $g$, it will not always grow at a constant rate (as AoI) in practice.  Hence,  we first consider hierarchical AoII processes, defined in our Definition \ref{definition}. That is to say, in this work,  the ageing process using AoII  depends on the system state, which can have a zero, or linear, or exponential growth, thus hierarchy, at different system states.
	
	To summarize, our main contributions in this work are: 
	
	$\bullet$ We give the closed-form expressions of average hierarchical AoII in two typical real-time scenarios: 1). A source delivers updates to a monitor over a noisy channel,  where the arrival and transmission of source updates are formulated as an M/M/1/1 queue; 2). A source of interest and a contender source both deliver updates to a monitor over a collision channel, where  the arrival and transmission of updates from each source are formulated as an M/M/1/1 queue, respectively. The system diagrams are depicted in Fig. \ref{system} and \ref{system2}. 
	
	$\bullet$  We provide a systematic way to analyze the continuous hierarchical AoII over unslotted real-time systems using SHSs, where two kinds of hierarchy schemes are discussed. The first is a hierarchy of zero and linear growth of AoII processes, where the linear growth of AoII has different slopes at different states. The second is a hierarchy of zero, linear, and exponential growth of AoII processes.
	
	$\bullet$ Following \cite{yates_age_2019} and \cite{maatouk_age-aware_2023}, we discuss the stability issue of hierarchical AoII under SHSs. We first prove a \textit{weak-sense stability}, i.e., positive recurrence, for the AoII SHSs in Lemma \ref{theorem11}. Then, we give the steady-state stability conditions in Theorem \ref{theorem2} and \ref{theorem3} for the two different scenarios  in this work, receptively, which  ensure the existence of  average AoII.
	
	$\bullet$ In Section \ref{Comparisons between AoI and AoII: The M/M/1/1 queue}, several classical results of  average AoI  under the M/M/1/1 queue model are compared with the average AoII in this work. A few insights are obtained based on the inherent differences between AoI and AoII: i). The ceasing of ageing using AoII sharply cuts down the average age in the M/M/1/1 queue compared to the results using AoI; ii). The average age using AoII can even be smaller than a generate-at-will queue with the same Poisson service rate $\mu$ for some content-change probability $p$; iii). The monotonicity of the average age with the normalized utilization factor $\rho=\frac{\lambda}{\mu}$ is different between these two age metrics.

	$\bullet$ The average AoII performance under different channel parameters  is evaluated. In the noisy channel case,  the average AoII grows monotonically with $p_e$ regardless of the utilization factor $\rho$ and content-changed probabilities $p$. In the collision channel case, the average AoII  grows monotonically with a larger $\rho$ since a higher arrival rate of updates from each source gives a higher collision probability. The content mismatch cannot be eliminated as soon as possible due to collisions, which  results in a larger average AoII.
	
	$\bullet$ The SHSs analysis  can be generalized to $M\geqslant3$ sources. The computation complexity using Theorem \ref{theorem1} to obtain more general results of multi-access networks is discussed in Section \ref{Discussions}, where the number of discrete states grows at least four times the number of nodes $M$.

	\section{System Model}\label{II}
	We first give our basic assumptions for the noisy channel case with single source and the collision channel case with two sources. Then we present the basic settings and relevant definitions for the AoII SHSs.
	
	\subsection{Model Assumptions}
	
	\subsubsection{The Splitting Poisson Process}
	
	We track the average AoII performance of a source node observing a specific information process $X(t)$. The node can be viewed as a sensor in IoT networks or an observing unit in vehicular networks.  The updates $\{X_t: t\in\mathbb{R}\}$ are sampled from $X(t)$ following a Poisson process with rate $\lambda$.   In this work, we consider a memoryless source model, where each newly-sampled update $X_t$ changes its content with probability $p$ and remain unchanged with probability $q=1-p$. The sampling (arriving) process at the transmitter therefore splits into two Poisson streams with rates $p\lambda$ and $q\lambda$. Moreover, following the discussion in \cite{meng_toward_2023, bountrogiannis_age_2025, liu_wireless_2023}, we  assume that the state space of the content of $X_t$ is very large (possibly infinite) and unknown, where the likelihood of content returning to a previous state is negligible if it has changed its state.
	
	The contender source  in the collision case is denoted as $X^c(t)$, whose updates $\{X^c_t:t\in\mathbb{R}\}$ are sampled by a node at a Poisson rate $\lambda_c$. We assume that the content of $X_t$ is independent of the content of $X^c_t$. Therefore, the AoII process of $X(t)$ measured at the monitor is only affected by $X^c_t$ in a sense that whether or not a collision arises.
	
	\subsubsection{Transmission Through Unreliable Channels}
	Since time is unslotted, a transmission immediately starts when a new update arrives. We  assume the transmission times of the updates are modeled as independent exponential $\mu$ and $\mu_c$ random variables for updates $X_t$ and $X^c_t$, respectively. Furthermore, the transmission of updates is preemptive considering that the content-changed updates should be delivered as soon as possible. Hence, we assume that the transmission of updates from $X(t)$ and $X^c(t)$ both follows M/M/1/1, where the preempted updates are discarded.
	
\begin{figure}[!t]
		\centering
		\scalebox{0.7}{
			\subfloat[]{\label{system}
				\begin{tikzpicture} 
					\coordinate (A) at (0,0); 
					\coordinate (B) at (4,0);     
					\draw[line width=1pt] (A) ellipse[x radius=0.75, y radius=0.75]; 
					\node[above] at (A)  {Source};
					\node[below] at (A)  {$X(t)$};
					\draw[-latex,thick] (0.75 ,0)--(1.7,0);
					\node [above] at (1.25,0) {$\lambda$};
					\node [rectangle, draw=black, thick] at (2,0) {$X_t$};
					\draw[-latex, thick] (2.32,0)--(3.25,0);
					\node [above] at (2.75,0) {$p_c\mu$};
					\draw[line width=1pt] (B) ellipse[x radius=0.75, y radius=0.75]; 
					\node[above] at (B) {Monitor};
					\node[below] at (B) {$\hat{X}_t$};
					\draw[->|, thick] (2.32,0)--(3,-0.5);
					\node [left] at (3.1, -0.7) {$p_e\mu$};
				\end{tikzpicture}
		}}
		\scalebox{0.7}{
			\subfloat[]{
				\label{system2}
				\begin{tikzpicture} 
					\coordinate (A) at (0,1); 
					\coordinate (B) at (4,0);  
					\coordinate (C) at (0,-1);    
					\draw[line width=1pt] (A) ellipse[x radius=0.75, y radius=0.75]; 
					\node[above] at (A)  {Source};
					\node[below] at (A)  {$X(t)$};
					\draw[-latex,thick] (0.75,1)--(1.7,1);
					\node [above] at (1.25,1) {$\lambda$};
					\node [rectangle, draw=black, thick] at (2,1) {$X_t$};
					\draw[line width=1pt] (C) ellipse[x radius=0.75, y radius=0.75]; 
					\node[above] at (C)  {Source};
					\node[below] at (C)  {$X^c(t)$};
					\draw[-latex,thick] (0.75,-1)--(1.7,-1);
					\node [above] at (1.25,-1) {$\lambda_c$};
					\node [rectangle, draw=black, thick] at (2,-1) {$X^c_t$};
					\draw[-latex, thick] (2.32,1)--(3.25,0);
					\node [above] at (2.75,0.7) {$\mu$};
					\draw[-latex, thick] (2.33,-1)--(3.25,0);
					\node [below] at (2.8, -0.7) {$\mu_c$};
					\draw[line width=1pt] (B) ellipse[x radius=0.75, y radius=0.75]; 
					\node[above] at (B) {Monitor};
					\node[below] at (B) {$\hat{X}_t$};
				\end{tikzpicture}
		}}
		\caption{Two system models with memoryless sources: (a) an M/M/1/1 queue over noisy channel. (b) two M/M/1/1 queues over collision channel.}
	\end{figure}
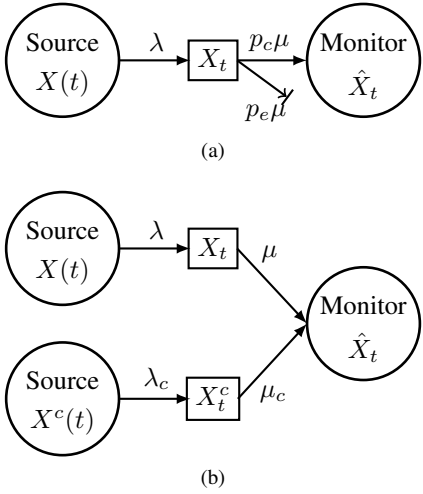

	The channel is unreliable in both cases. In Fig. \ref{system}, an update may be decoded incorrectly by the monitor due to channel noise. We denote $p_c$ as the probability of an update being correctly decoded, and an update that is not decoded correctly is lost with probability $p_e=1-p_c$. Note that only an update $X_t$ being correctly decoded can reduce the AoII at the monitor.
	
	In Fig. \ref{system2}, the channel is assumed to be noise-free yet suffering  collisions. A collision period begins if a newly sampled update $X^c_t$ / $X_t$ arrives while there is an update from $X_t$ / $X^c_t$ currently  being transmitted. The collision period ends if one of these updates finishes transmission. Hence, the collision disrupts transmission in the way that the update finishing transmission first cannot be successfully decoded and being discarded, whereas the remaining update would keep transmitting.  The  transmission time of the remaining is still exponential due to the memoryless property. This consideration leads to two different cases  that the remaining update can or cannot be successfully decoded,  both of which are addressed in Section \ref{IV}.
	
	\subsubsection{End-to-End Interaction}
	There are two common assumptions shared in real-time status-updating networks: i).  The amounts of communication overheads including feedback signals require negligible transmission time; ii). These overheads are transmitted via a dedicated control channel to ensure the signaling free from errors\footnote{To ensure reliability in such a control channel, a reliable protocol can be employed\cite{kurose_computer_2017}.}. Since  the state space of the content of $X_t$ is assumed to be unknown,  our source model in fact cannot require the  knowledge of source content, but rather the knowledge of a one-bit indicator of content changing at source. This can be implemented under locally-deployed processors, e.g.,  pre-trained neural networks,  at the source side that output such an indicator. Furthermore,  we  assume that the indicator of content-change is instantaneously known at the monitor based on i) and ii), which results in changes of system states. 
	\theoremstyle{remark}\newtheorem{remark00}[remarkcounter]{Remark}
	\begin{remark00} \label{remark00}
		Inspired by \cite{maatouk_age_2023}, we discuss a potential way to implement such an indicator, where  some locally-deployed processors, e.g.,  pre-trained neural networks,  at the source side are used to output the indicator.   Define $I_s= \mathbb{I}\{c(X_{t_1}, X_{t_0})\geqslant d\}$, where $\mathbb{I}\{\cdot\}$ is an indicator function taking values from 0 and 1. The function $c(.,.)$ is some cost function that measures the content mismatch between two updates,  and $d$ is some predefined threshold. Note that $c(.,.)$  can be designated or fitted by the pre-trained neural network.  The function $f$ determines whether a content change exists between $ X_{t_1}$ and the last sample $ X_{t_0}$. When $c(X_{t_1}, X_{t_0})\geqslant d$, $I_s=1$ triggering a content-change indicator to the monitor; otherwise, $I_s=0$. The threshold parameter $d$ controls the system's tolerance to content mismatches, which is adjustable for different sources. 
	\end{remark00}

	\subsection{Basic Settings of AoII SHSs}
	Following \cite{hespanha_model_2005}, SHSs are defined by a set of stochastic differential equations
	\begin{equation}\label{flows}
		\mathbf{\dot{x}}=\frac{\partial\mathbf{x}}{\partial t}= f^\prime(\mathbf{q}, \mathbf{x}, t) + g^\prime(\mathbf{q}, \mathbf{x},t)\frac{\partial\mathbf{z}}{\partial t}.
	\end{equation} 
	The system state $(\mathbf{q}(t), \mathbf{x}(t))$ is partitioned into the {\itshape discrete state} $ \mathbf{q}\in \mathbb{Q}$ consistent with the ordinary jump process, and a $\mathit{k}$-dimensional stochastic process $\mathbf{x}$ with piecewise continuous sample paths called the {\itshape continuous state}. Given a discrete state space $\mathbb{Q}$ and a $\mathit{k}$-dimensional vector $\mathbf{z}$ of independent Brownian motion processes, the continuous state $\mathbf{x}$ evolves with two state-dependent functions $f^\prime$ and $g^\prime$. Moreover, there is a family of {\itshape discrete transition} maps that tracks possible changes of the joint system state $(\mathbf{q}, \mathbf{x})$,
	\begin{equation}\label{mapping}
		(\mathbf{q}^+, \mathbf{x}^+) = \phi_{\mathit{l}}(\mathbf{q}, \mathbf{x}, t),\ \phi_{\mathit{l}}:\mathbb{Q} \times \mathbb{R}^n \times [0, \infty) \to \mathbb{Q} \times \mathbb{R}^n,  
	\end{equation}
	where $(\mathbf{q}^+, \mathbf{x}^+)$ denotes the joint state right after the transition map $\phi_{\mathit{l}}$ takes place, and $\mathit{l}\in\mathcal{L}=\{1, 2, ..., m\}$ denotes a set of transitions with transition intensities
	$\lambda_{\mathit{l}}(\mathbf{q}, \mathbf{x})$.
	
	Based on the general settings of SHSs above, the continuous state can be defined as the AoII processes of interest in our models where $\mathbf{x}(t) = [x_1(t), x_2(t)]$. We denote $x_2(t)$ as the AoII of $X_t$ at the monitor, and $x_1(t)$ is what the AoII at the monitor would become if an update in service were to complete transmission at time $t$.  Since the AoII evolves deterministically at each state $\mathbf{q}$, we have $g^\prime(\mathbf{q}, \mathbf{x},t) = 0$ and $\mathbf{z}=0$. Moreover, the transition rates $\lambda_l$ of the AoII SHSs are all constant Poisson rates according to the system model.
	
	Above all, the \textit{hierarchy} of AoII of $X_t$ is characterized by the state-dependent function $f^\prime(\mathbf{q}, \mathbf{x}, t)$, which is called the \textit{growth rate function} in our paper.  Since we adopt  SHSs to analyze our model using the continuous AoII \eqref{gAoII}, the growth rate function $f^\prime(\mathbf{q}, \mathbf{x}, t)$ can be interpreted as the derivative of AoII at different discrete state $\mathbf{q}$. For example, if we have the same content between the source and monitor at some state $\mathbf{q}_1$, then $f^\prime(\mathbf{q}_1, \mathbf{x}, t)=0$. However, if we use a constant growth rate at  some state $\mathbf{q}_2$ when there is a content-mismatch, then  $f^\prime(\mathbf{q}_2, \mathbf{x}, t)=m$ where $m$ is a positive constant. That is, the ageing process at $\mathbf{q}_2$ grows like AoI with slope $m$. Moreover, if we use a linear growth rate\footnote{The terminology  follows from \cite[Lemma 1]{hespanha_model_2005}} in $\mathbf{x}$ at some state $\mathbf{q_3}$ when the content-mismatch is considered to be severe, then  $f^\prime(\mathbf{q}_3, \mathbf{x}, t)=k\mathbf{x}$. Then, the ageing process  grows exponentially at $\mathbf{q_3}$. To that end, we can define the concept \textit{hierarchical AoII} used in this paper.
	\newtheorem{definition2}[defcounter]{Definition}
	\begin{definition2}[Hierarchical  AoII]\label{definition}
		Under SHSs model,	the hierarchical AoII using the continuous form \eqref{gAoII} has different growth rate functions $f^\prime(\mathbf{q}, \mathbf{x}, t)$ at different system states.
	\end{definition2}
	
	In order to reveal closed-form results of the average hierarchical AoII, we  adopt an indicator function $I=\mathbb{I}\{X_t\neq\hat{X}_t\}$ to specify the information penalty function $g$ in \eqref{gAoII}, which is a common consideration in the AoII literature.   The indicator function $I$  takes values from 0 or 1 if the content of update being transmitted is the same or not as the estimate $\hat{X}_t$, i.e., the last successfully decoded update, stored at the monitor. Hence, such an indicator is state-dependent in AoII SHSs, i.e., $I=I(\mathbf{q})$. We also note that, for the memoryless source with zero returning probability,  using an indicator of the content-change between two successive source updates is the same as using $I(\mathbf{q})$. Moreover, based on Definition \ref{definition}, the age penalty function $f$ that adopts to characterize hierarchical AoII can have a linear or exponential growth depending on  the choice of $f^\prime(\mathbf{q}, \mathbf{x}, t)$ for different $\mathbf{q}$. So, we can rewrite \eqref{gAoII} as  
		\begin{equation}\label{sAoII}
				x(\mathbf{q},\mathbf{x}, t) = f(\mathbf{q},\mathbf{x}, t)I(\mathbf{q}),\ t\in \mathbb{R}.
			\end{equation}
   
   Above all, since we leverage SHSs to formulate our models, in the following, we characterize the specified hierarchical AoII processes $x(\mathbf{q},\mathbf{x}, t)$ by adopting different hierarchical schemes of $f^\prime(\mathbf{q}, \mathbf{x}, t)$  under the considered scenarios.

	\subsection{Stability Analysis of SHSs - Fundamentals}
	In studying SHSs, researchers use Lagrange stability to depict systems that are ultimately uniformly bounded in the $n$-th mean ($n\geqslant1$) of $\mathbf{x}(t)$\cite{yin_hybrid_2010,teel_stability_2014,maatouk_age-aware_2023}. However,  as suggested in \cite[Chapter 3]{yin_hybrid_2010}, for systems that only focus on the time and ensemble average of $\mathbf{x}(t)$, it is more appropriate to  analyze a \textit{weak-sense stability}, i.e., positive recurrence, which encloses the case  $n=1$ in Lagrange stability. 
	
	\newtheorem{definition}[defcounter]{Definition}
	\begin{definition}[Recurrence and Positive Recurrence\cite{yin_hybrid_2010}]
		For any set $\mathbb{V}=\mathbb{Q}\times D\subset \mathbb{Q}\times \mathbb{R}^n$ and an initial condition $(\mathbf{q}_0, \mathbf{x}_0)\in\mathbb{V}^c$,  the complement of $\mathbb{V}$,  the recurrent time is the time interval that $(\mathbf{q}, \mathbf{x})$ at time $t$ returns to  $\mathbb{V}$ from the initial state, i.e.,
		\begin{equation}
			\sigma_{\mathbb{V}} = \inf\{t:(\mathbf{q}_0, \mathbf{x}_0)\in\mathbb{V}^c,(\mathbf{q}, \mathbf{x}) \in \mathbb{V}\},
		\end{equation}
		A regular SHS process is \textit{recurrent} with respect to $\mathbb{V}$ if
		\begin{equation}\label{recurrent}
			\mathrm{Pr}[\sigma_{\mathbb{V}} <\infty]=1,\ \forall\ (\mathbf{q}_0, \mathbf{x}_0)\in\mathbb{V}^c.
		\end{equation}
		In addition, a recurrent process with finite mean recurrence time for some $\mathbb{V}$ is \textit{positive  recurrent} with respect to $\mathbb{V}$.
	\end{definition}
	A regular SHS process is an SHS that always has a unique global solution for any initial condition. We assume that the regularity holds for the AoII SHSs from now on, which is verified later in the proof in Appendix \ref{app a1}.
	
	The Lagrange stability  of SHSs is discussed in depth by Maatouk et al. in \cite{maatouk_age-aware_2023}. They strictly prove the positive recurrence of the proposed age-aware SHSs and also prove the Lagrange stability for the higher-order age moments.  Following them, we  give our stability analysis of our AoII SHSs first from the perspective of positive recurrence since we are only interested in the average AoII. 
	
	In addition, we discuss the necessary conditions that are  determining factors to ensure the existence of steady-state (or ergodic) average AoII after we obtain the positive recurrence. The analysis of such conditions was first carried out by Yates in \cite[Section V]{yates_age_2019}.  Following this, we then analyze the  steady-state stability conditions  of the AoII SHSs based on ergodicity.

	\begin{figure}
		\centering
		\subfloat[]{\label{linear case}
			\begin{tikzpicture}
				[age2/.style={green!30!blue!50,very thick}, age3/.style={ dashed}]
				\draw[age2] (0,0)--(1,1)--(1.5,2);
				\draw[age3] (1.5,2)--(1.5,0);
				\draw[age3] (1,0)--(1,2);
				\draw[age3] (2.5,0)--(2.5,0.5);
				\draw[age3] (2,0)--(2,2);
				\draw[age3] (2.5,0)--(2.5,2);
				\draw[age2] (1.5,0)--(2,0)--(2.5,0.5)--(3, 2);
				
				\draw[-stealth,line width=0.75pt] (0,0) -- (3.5,0.0); 
				\draw[-stealth,line width=0.75pt] (0,0) -- (0.0,2.5);

				\node[right] at (3.5,0.0) {\footnotesize{$t$}};
				\node[above] at (0.0,2.5) {\footnotesize{$x_2(t)$}};
				
				\node[above] at (0.5, 1) {\footnotesize$1_m$};
				\node[above] at (2.3, 1) {\footnotesize$1_m$};
				\node[above] at (3, 2) {\footnotesize$1_u$};
				\node[above] at (1.3, 2) {\footnotesize$0_e$};
				\node[above] at (1.8, 1) {\footnotesize$0_c$};
				
				\coordinate (a) at (1.5,0);
				\fill (a) circle[radius=1.5pt]; 
				\node[below] (t0) at (a) {\footnotesize{$t_{0_c}$}};
			\end{tikzpicture}
		}
		\subfloat[]{\label{quadratic case}
			\begin{tikzpicture}
				[age2/.style={green!30!blue!50,very thick},  age3/.style={dashed}]
				\draw[age2] (0,0)--(1,1);
				\draw[age2] (1, 1) arc (-50:-10:2);
				\draw[age3] (1.7,2.2)--(1.7,0);
				\draw[age3] (1,0)--(1,2.2);
				\draw[age3] (2.2,0)--(2.2,2.2);
				\draw[age3] (2.5,0)--(2.5,2.2);
				\draw[age2] (1.7,0)--(2.2, 0)--(2.5, 0.3);
				\draw[age2] (2.5, 0.3) arc (-40:0:2.7);
				
				\draw[-stealth,line width=0.75pt] (0,0) -- (3.5, 0.0); 
				\draw[-stealth,line width=0.75pt] (0,0) -- (0.0, 2.5);
				
				\node[above] at (0.5, 1) {\footnotesize$1_m$};
				\node[above] at (2.35, 1) {\footnotesize$1_m$};
				\node[above] at (2, 1) {\footnotesize$0_c$};
				\node[above] at (3, 2.1) {\footnotesize$1_u$};
				\node[above] at (1.3, 2.1) {\footnotesize$0_e$};
				
				\node[right] at (3.5,0.0) {\footnotesize{$t$}};
				\node[above] at (0.0,2.5) {\footnotesize{$x_2(t)$}};
				
				\coordinate (a) at (1.7,0);
				\fill (a) circle[radius=1.5pt]; 
				\node[below] (t0) at (a) {\footnotesize{$t_{0_c}$}};
			\end{tikzpicture}
		}
		\caption{Sample paths of AoII $x_2(t)$ at monitor in the noisy channel case under: (a) hierarchy scheme 1; (b) hierarchy scheme 2.  $t_{0_c}$ is the time instance that the content is matched.} 
		\label{AoII}
	\end{figure}
	\section{SHSs Analysis over Noisy Channel: The M/M/1/1 queue revisited}\label{III}
	The AoI of the M/M/1/1 queue has been thoroughly studied under different queueing disciplines. e.g., \cite{costa_age_2016, najm_status_2018, yates_age_2020b}. However, the queue has never been examined using a context-aware hierarchical AoII (ageing) process. In this section, we provide a systematic way to obtain the closed-form results of the average hierarchical AoII in \eqref{sAoII} using SHSs. The results would be later compared with the average AoI in Section \ref{numerical}.

	\subsection{SHSs Formulations with Hierarchical AoII}\label{SHSs Formulations with Hierarchical AoII}
	To begin, we first need to characterize the system state, i.e., $\mathbf{q}\in\mathbb{Q}=\{0_c,0_e,1_m,1_u\}$, under the noisy channel case. State $0_c$ is defined as the correct (content-matched) state, where the AoII would not increase hence $\mathbf{\dot{x}}=[0,0]=\mathbf{0}_n$. State $1_m$ is defined as the mismatched state, where a content-changed update arrives and starts transmission. The AoII $\mathbf{x}(t)$ grows as the traditional AoI in this state, i.e., $\mathbf{\dot{x}}=[1,1]=\mathbf{1}_n$. State $0_e$ is defined as the error state, where no update is on transmission yet the content between ends is mismatched due to the channel noise. Hence, the age processes $\mathbf{x}(t)$ keep growing in $0_e$. State $1_u$ is the urgent state, where the transmitter is delivering a more important update. According to the model assumptions, $1_u$ can be reached when: 
	
	i). A newly sampled update with changed content preempts transmission. In that case, the content of  $\hat{X}_t$ lags at least two versions behind the transmitting update; 
	
	ii). A newly sampled update recovers the system from the error state. In that case, the monitor needs the update to fix the error as soon as possible.
	
	Compared to states $0_c$ and $1_m$, the level of age dissatisfaction at states $0_e$ and $1_u$ is considered to be severe for the following two reasons.
	
	$\bullet$ The content mismatch between ends cannot be directly eliminated from the error state $0_e$ since there is no update on transmission. Being away from $0_c$ compared to the other states exacerbates the ageing of $\hat{X}_t$ at the monitor.
	
	$\bullet$ The content of update being transmitting at $1_u$ is urged by the monitor due to issues i) and ii)  given above, which also exacerbates the ageing of $\hat{X}_t$. 
	
	Based on these considerations, the AoII processes at the monitor should not be the same under different system states, in other words, hierarchical AoII is needed.  To that end, we construct two different hierarchy schemes, each with four hierarchical levels, using the growth rate function $f^\prime$. 
	
	- \textit{Hierarchy Scheme 1}:
	\begin{equation}\label{linear function}
		f_1^\prime(\mathbf{q}, \mathbf{x}, t)=
		\begin{cases}
			\mathbf{0}_n, \quad \textrm{if}\ \mathbf{q}=0_c,\\
			\mathbf{1}_n,\quad  \textrm{if}\ \mathbf{q}=1_m,\\
			m_1\mathbf{1}_n,\quad \textrm{if}\ \mathbf{q}=0_e,\\
			m_2\mathbf{1}_n,\quad \textrm{if}\ \mathbf{q}=1_u.\\
		\end{cases}
	\end{equation}
	
	Scheme 1 represents the case that  $f^\prime$ has four levels of constant growth rates at each state. The parameters $m_2\geqslant m_1\geqslant1$ are slopes representing two different levels of age dissatisfaction in this case.
	
	- \textit{Hierarchy Scheme 2}:
	
	\begin{equation}\label{quadratic function}
		f_2^\prime(\mathbf{q}, \mathbf{x}, t)=
		\begin{cases}
			\mathbf{0}_n,\quad \textrm{if}\ \mathbf{q}=0_c,\\
			\mathbf{1}_n,\quad \textrm{if}\ \mathbf{q}=1_m,\\
			k_1\mathbf{x},\quad \textrm{if}\ \mathbf{q}=0_e,\\
			k_2\mathbf{x},\quad \textrm{if}\ \mathbf{q}=1_u.\\
		\end{cases}
	\end{equation}  
	
	Scheme 2 represents the case that  $f^\prime$ has two levels of constant growth rates and two levels of linear growth rates in $\mathbf{x}$. 
	The parameters $k_1$ and $k_2$, where $k_2\geqslant k_1$, are the growth constants controlling the stability of SHSs, which will later be used in obtaining the stability conditions of AoII SHSs.

	\theoremstyle{remark}\newtheorem{remark}[remarkcounter]{Remark}
	\begin{remark} \label{remark1}
		The subsequent results using \eqref{linear function} or \eqref{quadratic function} can be trivially simplified to a three-level hierarchical case by letting $m_1=m_2$ or $k_1=k_2$. In addition,  if we change the level of age dissatisfaction at state $1_u$ in scheme 2 into the level of  constant growth rate, the analysis of such AoII SHSs is basically the same as scheme 2. Therefore, we take \eqref{quadratic function} as a common example in this work. 
	\end{remark}

	\subsection{Generator, Positive Recurrence and Calculation of SHSs}
	To quantify the average AoII, we first need to employ a test function to track the dynamics of AoII at each state $\overline{\mathbf{q}}\in\mathbb{Q}$,
	\begin{equation}\label{test function}
		\psi_{\overline{\mathbf{q}}}(\mathbf{q},\mathbf{x},t)=\mathbf{x}(t)\delta_{\overline{\mathbf{q}},\mathbf{q}},
	\end{equation}
	where $\delta_{\overline{\mathbf{q}},\mathbf{q}}$ is the Kronecker delta function that equals to one only if $\overline{\mathbf{q}}=\mathbf{q}$, otherwise it is zero. We also define 
	\begin{equation}\label{test}
		\mathrm{E}[\psi_{\overline{\mathbf{q}}}(\mathbf{q},\mathbf{x},t)]=\mathrm{E}[\mathbf{x}(t)\delta_{\overline{\mathbf{q}},\mathbf{q}}]=\mathbf{v}_{\overline{\mathbf{q}}}(t),
	\end{equation}
	as the expected value of AoII in each state $\overline{\mathbf{q}}$ at time $t$, where $\mathbf{v}_{\overline{\mathbf{q}}}(t)=[v^1_{\overline{\mathbf{q}}}(t),v^2_{\overline{\mathbf{q}}}(t)]$ tracks two different AoII process. Furthermore, we  define the stationary solution (if exists) when $t\to\infty$ to be $\mathbf{v}_{\overline{\mathbf{q}}}=[v^1_{\overline{\mathbf{q}}},v^2_{\overline{\mathbf{q}}}]=\lim\nolimits_{t\to\infty}\mathbf{v}_{\overline{\mathbf{q}}}(t)$.
	
	Given that $\psi$ is specified and $g(\mathbf{q}, \mathbf{x},t) = 0$, we introduce an operator $L$ on the test function $\psi$\cite{hespanha_modelling_2006}, that is,
	
	\begin{equation}\label{generator}
		\small
		\begin{aligned}
			(L\psi)(\mathbf{q}, \mathbf{x}, t)
			=&f^\prime(\mathbf{q}, \mathbf{x}, t)\frac{\partial\psi(\mathbf{q}, \mathbf{x}, t)}{\partial\mathbf{x}}\\
			&+\sum_{\mathit{l}=1}^{m}\lambda_{\mathit{l}}\Big[\psi(\phi_{\mathit{l}}(\mathbf{q}, \mathbf{x}, t), t) - \psi(\mathbf{q}, \mathbf{x}, t)\Big],
		\end{aligned}
	\end{equation} 
	which is called the extended generator for our SHSs. $f^\prime(\mathbf{q}, \mathbf{x}, t)$ is in the form of \eqref{linear function} or \eqref{quadratic function}, and $\lambda_l$ for the noisy channel case is given in Table \ref{table1}.  Note that the extended generator  $(L\psi)_{0_c}=(L\psi)_{0_c}(\mathbf{q}, \mathbf{x}, t)$ at state $0_c$ is zero since the AoII at $0_c$ never grows at all.\footnote{ As noted in \cite{yates_age_2019}, $0_c$  is called the irrelevant state to AoII. However, this kind of states does affect the stationary distributions of discrete state $\mathbf{q}$.}
	
	With these quantities defined, we are able to introduce the main tool of SHSs in Lemma \ref{main}, known as \textit{Dynkin formula}.
	\theoremstyle{plain}\newtheorem{generator}[lemmacounter]{Lemma}
	\begin{generator}\label{main}
		Supposing there is a bounded measurable test function $\psi:\mathbb{Q}\times\mathbb{R}^n\times[0, \infty)\to\mathbb{R}$, we have
		\begin{equation}\label{diff dynkin}
			\frac{\mathrm{d}\mathrm{E}[\psi(\mathbf{q}, \mathbf{x}, t)]}{\mathrm{d}t} = \mathrm{E}[(L\psi)(\mathbf{q}, \mathbf{x}, t)].
		\end{equation}
	\end{generator}
	
	Lemma \ref{main} in fact tells that the dynamics of the test function are on average characterized by its extended generator $L\psi$, whose expected value is the expected rate of $\psi$.

	\subsubsection*{Positive Recurrence}
	To move forward, we need to establish the weak-sense stability, i.e., positive recurrence, for our AoII SHSs under both hierarchy schemes.
	
	\theoremstyle{plain}\newtheorem{pr}[lemmacounter]{Lemma}
	\begin{pr}\label{theorem11}
		The AoII SHSs under the noisy channel case are positive recurrent in both hierarchy schemes.
	\end{pr} 
	
	The proof of Lemma \ref{theorem11} appears in Appendix \ref{app a1}, which can also be used in proving the positive recurrence of AoII SHSs under the collision channel case. Lemma \ref{theorem11} also guarantees that the AoII SHSs are regular under both hierarchy schemes where the expected AoII $\mathrm{E}[\mathbf{x}(t)]$ is always finite with probability one. Moreover,  we can establish the ergodicity of the AoII SHSs based on the positive recurrence using the same technique introduced in \cite{maatouk_age-aware_2023} to prove steady-state convergence of average AoII\footnote{The technique of proof has been revealed in our work \cite[Theorem 2]{90}, hence it is omitted here  for simplicity.}. With these, we can now proceed with our calculation for closed-form expressions of average AoII for the M/M/1/1 queue under  $f_1^\prime(\mathbf{q})$, which basically follows \cite[Theorem 1]{yates_age_2020b}. However, in order to calculate the average AoII under $f_2^\prime(\mathbf{q})$, we need to modify the aforementioned theorem. Hence, we summarize our first result in the following.
	
	\theoremstyle{plain}\newtheorem{ergodicity}[theoremcounter]{Theorem}
	\begin{ergodicity}\label{theorem1}
		Define two sets of transitions
		\begin{equation}\label{two transition sets}
			\begin{aligned}
				\mathcal{L}^i_{\overline{\mathbf{q}}}=\{l\in\mathcal{L}:\overline{\mathbf{q}}\ \textrm{is the incoming state} \},\\
				\mathcal{L}^o_{\overline{\mathbf{q}}}=\{l\in\mathcal{L}:\overline{\mathbf{q}}\ \textrm{is the outgoing state}\},\\
			\end{aligned}
		\end{equation}
		as the respective sets of incoming and outgoing transitions for each state $\overline{\mathbf{q}}\in\mathbb{Q}$ associated with $l$. Then, we have:
		
		(i). The  stationary distributions $\{\pi_{\overline{\mathbf{q}}}:\overline{\mathbf{q}}\in\mathbb{Q}\}$ of discrete states in SHSs are given by the global balance equations:
		\begin{equation}\label{global balance equation}
			\pi_{\overline{\mathbf{q}}}\sum_{l\in\mathcal{L}^o_{\overline{\mathbf{q}}}}\lambda_l=\sum_{l\in\mathcal{L}^i_{\overline{\mathbf{q}}}}\lambda_l\pi_{\mathbf{q}_l}.
		\end{equation}
		
		(ii). If the growth rate function $f^\prime$ is constant at state $\overline{\mathbf{q}}$, the expected value $\mathbf{v}_{\overline{\mathbf{q}}}$ can be calculated by 
		\begin{equation}\label{expect linear}
			\mathbf{v}_{\overline{\mathbf{q}}}\sum_{l\in\mathcal{L}^o_{\overline{\mathbf{q}}}}\lambda_l=	f^\prime(\overline{\mathbf{q}})\pi_{\overline{\mathbf{q}}}+\sum_{l\in\mathcal{L}^i_{\overline{\mathbf{q}}}}\lambda_l\mathbf{v}_{\mathbf{q}_l}\mathbf{A}_l,
		\end{equation}
		where $\mathbf{q}_l$ is denoted as the other state in $\mathcal{L}^i_{\overline{\mathbf{q}}}$. 
		
		(iii). If the growth rate function $f^\prime$ is linear in $\mathbf{x}$ at state $\overline{\mathbf{q}}$, the expected value $\mathbf{v}_{\overline{\mathbf{q}}}$  can be calculated by 
		\begin{equation}\label{expect quadratic}
			\mathbf{v}_{\overline{\mathbf{q}}}\Big[\sum_{l\in\mathcal{L}^o_{\overline{\mathbf{q}}}}\lambda_l-k_{\overline{\mathbf{q}}}\Big]=\sum_{l\in\mathcal{L}^i_{\overline{\mathbf{q}}}}\lambda_l\mathbf{v}_{\mathbf{q}_l}\mathbf{A}_l,
		\end{equation} 
		where  $k_{\overline{\mathbf{q}}}$ is the growth constant associated with $\overline{\mathbf{q}}$.
		
		Consequently, the closed-form expressions of average AoII at the monitor can be obtained by the summation of expected values at each state, i.e.,
		\begin{equation}\label{average AoII}
			x=\sum_{\overline{\mathbf{q}}\in\mathbb{Q}}v_{\overline{\mathbf{q}}}^2.
		\end{equation}
	\end{ergodicity} 
	
	The matrix $\mathbf{A}_l$ in \eqref{expect linear} and \eqref{expect quadratic} is the transition matrix associated with the linear mapping $\phi_{\mathit{l}}$. Since the first two parts of Theorem \ref{theorem1} have been proved in \cite{yates_age_2020b}, we omit here for simplicity. The proof of (iii) appears in Appendix \ref{app a}. Note that \eqref{expect quadratic} is only valid under some constrained parameters $k_\mathbf{q}$, which, in this work,  is called the steady-state stability conditions of the AoII SHSs and is discussed in Section \ref{Average AoII under Quadratic Hierarchy}.
	
	\theoremstyle{remark}\newtheorem{remark2}[remarkcounter]{Remark}
	\begin{remark2} \label{remark2}
		Theorem \ref{theorem1} gives the general procedure to calculate the average AoII for different hierarchy schemes. First, we need to calculate the stationary distribution for each finite-state SHSs Markov chains using \eqref{global balance equation}, the existence of which ensures the ergodicity. Then, the average AoII can be calculated separately at each state $\mathbf{q}$ using different state-dependent growth function $f^\prime(\mathbf{q})$, as derived in \eqref{expect linear} and \eqref{expect quadratic}. Hence, our theorem can be easily generalized to multi-level hierarchy of state-dependent AoII using any hybrid of constant and linear (in $\mathbf{x}$) growth rate functions. 
	\end{remark2}

	\subsection{Average AoII under Hierarchy Scheme 1}
	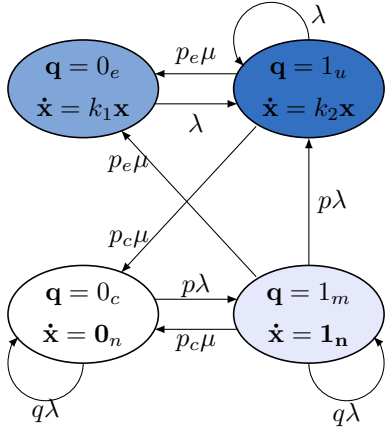
\begin{figure}
		\centering
		\scalebox{0.65}{
			\begin{tikzpicture}
				\coordinate (a) at (1.5,1.5);
				\filldraw[fill=green!30!blue!80, line width=0.6pt] (a) ellipse[x radius=1, y radius=0.65]; 
				\node [above] at (a) {$\mathbf{q}=1_u$};
				\node [below] at (a) {$\mathbf{\dot{x}} =  k_2\mathbf{x}$};
				\draw[-latex] (1.5,-0.85)--(1.5,0.85);
				\node[right] at (1.5, 0) {$p\lambda$};
				%			\draw[-latex] (1.7,0.85)--(1.7,-0.9);
				%				\node[right] at (1.6, 0) {$q\lambda$};
				\draw[-latex]  (0.56,1.7) -- (-0.56, 1.7);
				\node [above] at (0, 1.7) {$p_e\mu$};
				\draw[-latex] (-0.56, 1.3)-- (0.56, 1.3);
				\node[below] at(0, 1.3) {$\lambda$};
				\draw[-latex] (1.5,2.15) arc(0:225:0.5);
				\node [right] at (1.4, 2.5) {$\lambda$};
				\draw[-latex]  (0.8, 1)-- (-1, -0.95);
				\node[left] at(-0.5,-0.5) {$p_c\mu$};
				
				\coordinate (b) at (-1.5,1.5);
				\filldraw[fill=green!30!blue!50, line width=0.6pt] (b) ellipse[x radius=1, y radius=0.65];

				\node [above] at (b) {$\mathbf{q}=0_e$};
				\node [below] at (b) {$\mathbf{\dot{x}} = k_1\mathbf{x}$};
				
				\coordinate (c) at (1.5,-1.5);
				\filldraw[fill=green!10!blue!10,line width=0.6pt] (c) ellipse[x radius=1, y radius=0.65];
				\draw[-latex] (1.5,-2.15) arc(180:405:0.5);
				\node [below] at (2, -2.55) {$q\lambda$};
				\draw[-latex]  (0.8, -1)-- (-1, 0.95);
				\node[left] at(-0.5,0.5) {$p_e\mu$};
				%			\draw[-latex]  (-0.75, 1.1)-- (0.95, -1);
				%				\node[right] at(0.6,-0.5) {$q\lambda$};
				\node [above] at (c) {$\mathbf{q}=1_m$};
				\node [below] at (c) {$\mathbf{\dot{x}} = \mathbf{1_n}$};
				
				\coordinate (d) at (-1.5,-1.5);
				\coordinate (e) at (-0.56,-1.3);
				\draw[line width=0.6pt] (d) ellipse[x radius=1, y radius=0.65]; 
				\draw[-latex] (-1.5,-2.15) arc(0:-225:0.5);
				\node [below] at (-2, -2.5) {$q\lambda$};
				\draw[-latex] (e) -- (0.56,-1.3);
				\draw[-latex]  (0.56,-1.7) -- (-0.56, -1.7);
				\node [above] at (0, -1.4) {$p\lambda$};
				\node [below] at (0, -1.65) {$p_c\mu$};
				\node [above] at (d) {$\mathbf{q}=0_c$};
				\node [below] at (d) {$\mathbf{\dot{x}} = \mathbf{0}_n$};
		\end{tikzpicture}}		
		\caption{Graphical representation of SHSs: the M/M/1/1 queue with preemption over noisy channel. The hierarchical growth of AoII takes the second case as an example, which is distinguished by the color of state.}
		\label{SHS state}
	\end{figure}
	We now explain each transition $l$ with a constant $\lambda_{\mathit{l}}(\mathbf{q}, \mathbf{x})=\lambda_l$ under the noisy channel, which is illustrated in Fig. \ref{SHS state} and summarized in Table \ref{table1}  with the pre-defined SHSs states.
	\begin{table}[htbp]
		\caption{State Transitions and Changes of AoII over Noisy Channel\label{table1}}
		\centering
		\renewcommand{\arraystretch}{0.65}
		\scalebox{0.6}{
			\begin{tabular}{cccccc}
				\toprule[1pt]
				$l$ & $\mathbf{q}_l\to \mathbf{q}_l^+$ & $\lambda_l$  & $\mathbf{x}\mathbf{A}_l$ & $\mathbf{A}_l$	& $\mathbf{v}_{q_l}\mathbf{A}_l$ \\
				\midrule
				$0$ & $0_c\to 0_c$ & $q\lambda$  & $\begin{bmatrix} 0 & 0 \end{bmatrix}$ & $\begin{bmatrix} 0 & 0 \\ 0 & 0 \end{bmatrix}$ &  $ \begin{bmatrix} 0 & 0 \end{bmatrix}$ \\
				
				$1$ & $0_c\to 1_m$ & $p\lambda$  & $\begin{bmatrix} 0 & x_2 \end{bmatrix}$ & $\begin{bmatrix} 0 & 0 \\ 0 & 1 \end{bmatrix}$ &  $ \begin{bmatrix} 0 & v^2_{0_c} \end{bmatrix}$ \\
				
				$2$ & $1_m\to 0_c$ & $p_c\mu$  & $\begin{bmatrix} 0 & 0 \end{bmatrix}$ & $\begin{bmatrix} 0 & 0 \\ 0 & 0 \end{bmatrix}$ &  $ \begin{bmatrix} 0 & 0 \end{bmatrix}$ \\
				
				$3$ & $1_m\to 1_m$ & $q\lambda$  & $\begin{bmatrix} 0 & x_2 \end{bmatrix}$ & $\begin{bmatrix} 0 & 0 \\ 0 & 1 \end{bmatrix}$ &  $ \begin{bmatrix} 0 & v^2_{1_m}\end{bmatrix}$ \\
				
				$4$ & $1_m\to 0_e$ & $p_e\mu$  & $\begin{bmatrix} x_2 & x_2 \end{bmatrix}$ & $\begin{bmatrix} 0 & 0 \\ 1 & 1 \end{bmatrix}$ &  $ \begin{bmatrix} v^2_{1_m} & v^2_{1_m} \end{bmatrix}$ \\
				
				$5$ & $1_m\to 1_u$ & $p\lambda$  & $\begin{bmatrix} 0 & x_2 \end{bmatrix}$ & $\begin{bmatrix} 0 & 0 \\ 0 & 1 \end{bmatrix}$ &  $ \begin{bmatrix} 0 & v^2_{1_m} \end{bmatrix}$ \\
				
				$6$ & $1_u\to 0_c$ & $p_c\mu$  & $\begin{bmatrix} 0 & 0 \end{bmatrix}$ & $\begin{bmatrix} 0 & 0 \\ 0 & 0 \end{bmatrix}$ &  $ \begin{bmatrix} 0 & 0 \end{bmatrix}$ \\
				
				$7$ & $1_u\to 1_u$ & $\lambda$  & $\begin{bmatrix} 0 & x_2 \end{bmatrix}$ & $\begin{bmatrix} 0 & 0 \\ 0 & 1 \end{bmatrix}$ &  $ \begin{bmatrix} 0 & v^2_{1_u} \end{bmatrix}$ \\
				
				$8$ & $1_u\to 0_e$ & $p_e\mu$  & $\begin{bmatrix} x_2 & x_2 \end{bmatrix}$ & $\begin{bmatrix} 0 & 0 \\ 1 & 1 \end{bmatrix}$ &  $ \begin{bmatrix} v^2_{1_u} & v^2_{1_u} \end{bmatrix}$ \\
				
				$9$ & $0_e\to 1_u$ & $\lambda$  & $\begin{bmatrix} 0 & x_2 \end{bmatrix}$ & $\begin{bmatrix} 0 & 0 \\ 0 & 1 \end{bmatrix}$ &  $ \begin{bmatrix} 0 & v^2_{0_e} \end{bmatrix}$ \\
				
				%	$9$ & $1^+\to 1^-$ & $q\lambda$  & $\begin{bmatrix} 0 & x_2 \end{bmatrix}$ & $\begin{bmatrix} 0 & 0 \\ 0 & 1 \end{bmatrix}$ &  $ \begin{bmatrix} 0 & 0 \end{bmatrix}$ \\
				%	
				%	$9$ & $0^+\to 1^-$ & $q\lambda$  & $\begin{bmatrix} 0 & x_2 \end{bmatrix}$ & $\begin{bmatrix} 0 & 0 \\ 0 & 1 \end{bmatrix}$ &  $ \begin{bmatrix} 0 & 0 \end{bmatrix}$ \\
				
				\bottomrule[1pt]
			\end{tabular}
		}
	\end{table}
	
	$l=0,3,7$: Since the content of a newly sampled update remains unchanged with probability $q$ in the self-transition at state $0_c$ ($l=0$), we assume that this kind of update would be maintained at the transmitter subject to an exponential random interval with rate $\mu$ (then dropped) instead of being transmitted. In the self-transition at state $1_m$ ($l=3$), the preemption of an unchanged update also has no effect on the AoII. Moreover, since we consider the four-level hierarchy in the measure of timeliness for the noisy channel case,  the state $1_u$ would stay under any preemption while the AoII process reaches the fourth level, which is the highest, at $1_u$ ($l=7$).
	
	$l=1,5$: An update with changed content leads to the second-level dissatisfaction between ends and the AoII starts to grow after the state is transferred into $1_m$  ($l=1$). Similarly, the update under transmission would also be preempted (then discarded) by a newly sampled update with changed content, which leads to the fourth-level dissatisfaction after the state is transferred into $1_u$ ($l=5$).
	
	$l=2, 6$: The content between ends is matched after a successful transmission over the noisy channel. Meanwhile, the AoII is reset to zero by these two transitions, respectively.
	
	$l=4,8$:  If the update is not successfully decoded by the monitor due to noise, the system state would be transferred into the error state $0_e$. Then, the system suffers a severe age dissatisfaction, namely, the third-level in our consideration, due to the mismatch of content.
	
	$l=9$: If the system were in the error state $0_e$, any newly sampled update $X_t$ transfers the system state into $1_u$ and starts a transmission.

	Following Theorem \ref{theorem1}, the stationary distributions of the SHSs Markov chain in Fig. \ref{SHS state} are calculated as:
	\begin{equation}\label{linear stationary distribution}
		\begin{aligned}
			&\pi_{0_c}=\frac{p_c}{a},\ &\pi_{1_m}&=\frac{pp_c}{a(p+\rho^{-1})},\\ 
			&\pi_{0_e}=\frac{pp_e}{a},\  &\pi_{1_u}&=\frac{p^2\rho+pp_e}{a(p+\rho^{-1})},
		\end{aligned}
	\end{equation}
	where  $a = p+p\rho+p_cq$, and $\rho=\lambda/\mu$ is known as the utilization factor of the queue. Then we can use \eqref{expect linear} and \eqref{average AoII} to obtain the following results.
	
	\theoremstyle{plain}\newtheorem{result}[corollarycounter]{Corollary}
	\begin{result}\label{corollary1}
		For  hierarchy scheme 1 in \eqref{linear function}, the average AoII $x_a$ of a preemptive M/M/1/1 queue over the noisy channel is
		\begin{equation}\label{linear average AoII}
			x_a=\frac{m_1pp_e(1+\rho)}{a\lambda p_c}+\frac{(p_e+\rho)\big[pp_c+m_2(p^2\rho+pp_e)\big]}{a\lambda p_c(p+\rho^{-1})}.
		\end{equation}
		Furthermore, if we set $m_1=m_2=1$, then \eqref{linear average AoII} reduces to the simplest expression with only two levels of hierarchy,
		\begin{equation}\label{simplified linear AoII}
			x^\prime_a=\frac{p(p_e\rho^{-1}+2p_e+\rho)}{ap_c\mu}.
		\end{equation}
	\end{result}
	The derivation of \eqref{linear average AoII} appears in Appendix \ref{app b}.  The ergodicity of the SHSs Markov chain is established by \eqref{linear stationary distribution} and Lemma \ref{theorem11}. Hence, it is easy to verify that \eqref{linear average AoII} always exists for any finite $m_1$ and $m_2$ based on the invertibility of matrix $\mathbf{B}$ defined in \eqref{matrix1}.\footnote{A general proof can be found in \cite[Section V]{yates_age_2019}.} The simplified version \eqref{simplified linear AoII} would later be used to compare with the AoI results in Section \ref{numerical}. 
	
	Moreover, we note that \eqref{linear average AoII} and \eqref{simplified linear AoII} can be further specified to the results under metric AoS proposed in \cite{zhong_two_2018} by setting $p=1$.  The AoS metric is defined as the time since the content between two ends are out-of-sync. This is exactly the case of the proposed AoII metric where every newly sampled update changes its content with probability one. Hence, the expressions of average AoS in a preemptive M/M/1/1 queue over the noisy channel are given by
	\begin{equation}
		\begin{aligned}
			x_{a,AoS}&=\frac{m_1p_e}{\lambda p_c}+\frac{p_c(p_e+\rho)+m_2(p_e+\rho)^2}{\lambda p_c(1+\rho)(1+\rho^{-1})},\\
			x^\prime_{a,AoS}&=\frac{p_e\rho^{-1}+2p_e+\rho}{p_c\mu(1+\rho)},\ \textrm{for}\ m_1=m_2.
		\end{aligned}
	\end{equation}

	\subsection{Average AoII under Hierarchy Scheme 2}\label{Average AoII under Quadratic Hierarchy}
	The average AoII under scheme 2 is more complex since it involves two kinds of ageing patterns.  By leveraging Theorem \ref{theorem1}, the closed-form expression is summarized below.
	\theoremstyle{plain}\newtheorem{result2}[corollarycounter]{Corollary}
	\begin{result2}\label{corollary2}
		For hierarchy scheme 2 in \eqref{quadratic function}, the average AoII $x_b$ of a preemptive M/M/1/1 queue over the noisy channel is
		\begin{equation}\label{quadratic average AoII}
			x_b=\frac{pp_c(p_e\mu+\lambda-k_1)(p\lambda+\mu-k_2)}{a\lambda(p+\rho^{-1})^2\big[(\lambda-k_1)(\mu-k_2)-\lambda\mu p_e\big]},
		\end{equation}
		given that the  steady-state stability conditions are satisfied.
	\end{result2}
	
	The derivation of \eqref{quadratic average AoII} appears in Appendix \ref{app c}. However, we are more interested in the steady-state stability conditions of \eqref{quadratic average AoII}, where the invertibility of matrix $\mathbf{C}$ defined in \eqref{matrix2} depends on the values of $k_1$ and $k_2$. These conditions are given below, the proof of which can be found in Appendix \ref{app d}.
	
	\theoremstyle{plain}\newtheorem{stability}[theoremcounter]{Theorem}
	\begin{stability}\label{theorem2}
		The average hierarchical AoII in \eqref{quadratic average AoII} exists if and only if the following two conditions are all satisfied:
		
		i). The determinant of the associated matrix $\mathbf{C}$ is large than zero, i.e., $\det(\mathbf{C})=1-\frac{p_e\lambda\mu}{(k_1-\lambda)(k_2-\mu)}>0$;
		
		ii). The growth constants $k_1$ and $k_2$  are limited, i.e., $0<k_1<\lambda<\mu$ and $0<k_1<k_2<\mu$.
	\end{stability}
	
	Theorem \ref{theorem2} can be extended to other multi-level hierarchy scheme with finer state definitions as long as the hierarchy only consists of constant and linear growth rate functions. For example, one can split the state $0_e$ into two states $0^1_e$ and $0^2_e$ representing different levels of error, where $0^1_e$ is defined the same as the original error state directly connected to $1_m$ and $0^2_e$ is defined as a more severe error state that can only be reached by $1_u$. We also note that the above conditions are actually sufficient and necessary,  as proved in Appendix \ref{app d}.

	\section{SHSs Analysis over Collision Channel}\label{IV}
	In this section, we focus on another kind of unreliable channel that collides when multiple nodes attempt to use the channel simultaneously, in other words, the transmission time of these updates overlaps.  The SHSs analysis is basically the same as the noisy channel case, except that the system state space $\mathbb{Q}_c$ is larger resulting in a much more complex expression of the average AoII. Therefore,  under the collision channel, we only consider the hierarchy scheme with constant and linear growth rate functions since it is more representative. 
	
	In the following, we consider that only one contender source attempts to access the channel against the source of interest. The collision period ends if one of the updates finishes transmission first, which cannot be decoded and is discarded. Given that, there exists two different cases handling the remaining update on transmission: 1). It can be successfully decoded as long as there is no collision; 2). It cannot be successfully decoded.  There is no fundamental difference between the analysis of case 1) and 2) by SHSs, except that the state spaces and transitions for two cases are slightly different. Therefore, we first analyze case 1) in the main body of this paper, while the discussion of case 2) is provided in the supplementary material due to the limitation of space.
	
	\theoremstyle{remark}\newtheorem{remark0}[remarkcounter]{Remark}
	\begin{remark0} \label{remark0}
		In practice, the model of collision 1) is suitable for the situation that the lengths, i.e., the transmission time, of two collided updates are significantly different. The collision period (overlap) is relatively short compared to the transmission time of the update with large transmission time.  As a result, the effect of collisions on the content of remaining update is considered to be as small as possible. So the monitor is able to decode such updates under some advanced channel coding/decoding techniques. On the contrary, the model of collision 2) is suitable for the situation that the transmission time of two collided updates is close, which follows the collision model considered in \cite{yates_age_2020a}. Furthermore, since we allow preemption of updates for both sources, we assume that cases 1) and 2) are valid under preemption to facilitate the definition of discrete states in AoII SHSs. This assumption is only for the purpose of reducing the computation complexity of SHSs and demonstrating this work more clearly. 
	\end{remark0}
	
	\subsection{SHSs Formulations}
	For collision case 1), we first redefine the system states where the state space $\mathbb{Q}_c$ consists of eight different states $\mathbf{q}\in\mathbb{Q}_c=\{0_c,0_e,1_c,1_m,1_u,1_e,2_m,2_u\}$. States $0_c, 0_e, 1_m$ and $1_u$ are defined the same as the noisy channel case. However,  we have to added four more states due to the existence of $X^c_t$, namely, $1_c, 1_e, 2_m, 2_u$. State $1_c$ is defined as the other correct (content-matched) state while the contender  utilizes the channel to transmit $X^c_t$ alone. Hence, the AoII in $1_c$ would also not increase as in $0_c$. On the contrary, state $1_e$ is defined as  the other error state while the contender utilizes the channel yet the content between ends is mismatched, where we assume that the AoII in $1_e$ grows the same as in $0_e$. States $2_m$ and $2_u$ are defined as two distinct collision states with different AoII processes when the transmission of $X_t$ and $X^c_t$ overlaps. State $2_m$ can be directly reached by $1_m$ and $1_c$, which means the content of $\hat{X}_t$ at the monitor only lags one version behind that of $X_t$. However, the content of $X_t$ in $2_u$ is considered to be more urgent for the same two reasons i) and ii) given in \ref{SHSs Formulations with Hierarchical AoII}, so the AoII process of  $\hat{X}_t$ is more severe than $2_m$.
	
	Moreover, since we are only interested in the AoII behavior of $X_t$, the continuous states of our SHSs are defined the same as the noisy channel case, i.e., $\mathbf{x}(t) = [x_1(t), x_2(t)]$ due to the independence between $X_t$ and $X^c_t$.
	
	Based on the above state definitions, the hierarchy scheme considered here also has four levels, which is characterized by the state-dependent growth rate function $f_3^\prime(\mathbf{q}, \mathbf{x}, t)$.
	
	- \textit{Hierarchy Scheme 3}:
	
	\begin{equation}\label{quadratic function2}
		f_3^\prime(\mathbf{q}, \mathbf{x}, t)=
		\begin{cases}
			\mathbf{0}_n,\quad \textrm{if}\ \mathbf{q}=0_c,1_c,\\
			\mathbf{1}_n,\quad \textrm{if}\ \mathbf{q}=1_m, 2_m,\\
			k_1\mathbf{x},\quad \textrm{if}\ \mathbf{q}=0_e, 1_e, 1_u,\\
			k_2\mathbf{x},\quad \textrm{if}\ \mathbf{q}=2_u.\\
		\end{cases}
	\end{equation}  
	
	With a slight notation abuse, the parameters $k_1$ and $k_2$ are also denoted as the growth constants controlling the steady-state stability of SHSs, where $k_2\geqslant k_1$.  Given that the collision channel is noise-free, state $2_u$ is assumed to be the only urgent state with the highest age dissatisfaction (the fourth level)  since it is the only state that suffers collisions and also a severe content-mismatch. Note that this assumption can be easily generalized by adapting the growth rate constant $k_1$ and $k_2$.
	\theoremstyle{remark}\newtheorem{remark3}[remarkcounter]{Remark}
	\begin{remark3} \label{remark3}
		As we emphasized in the noisy channel case, the analysis in this section can be extended to other multi-level hierarchy schemes using finer state definitions. For example, one can track different urgent states in which successive arrivals of content-changed updates can lead to different levels of age dissatisfaction. These extensions are not trivial even in the two node case. To that end, each hierarchy scheme needs to be carefully designed since the increasing of the number $M$ using finer  state definitions makes the inverse of matrix in the AoII SHSs (very) hard to obtain, as shown in the case below.
	\end{remark3}
	\begin{figure}[htbp]
		\centering
		\scalebox{0.6}{
			\begin{tikzpicture}
				\coordinate (a) at (0,1);
				\filldraw[fill=green!30!blue!50, line width=0.6pt] (a) ellipse[x radius=1, y radius=0.65]; 
				\node [above] at (a) {$\mathbf{q}=1_u$};
				\node [below] at (a) {$\mathbf{\dot{x}} = k_1\mathbf{x}$};
				\draw[-latex] (0,1.65) arc(0:225:0.5);
				\node [below] at (-0.6, 2.2) {$\lambda$};
				\draw[-latex]  (0.92, 1.2) -- (2.08, 1.2);
				\node [above] at (1.5, 1.1) {$\lambda_c$};
				\draw[-latex]  (-0.8, 0.6) -- (-2.2, -0.6);
				\node [left] at (-1.7, -0.2) {$\mu$};
				
				\coordinate (b) at (0,-1);
				\filldraw[fill=green!10!blue!10, line width=0.6pt] (b) ellipse[x radius=1, y radius=0.65]; 
				\node [above] at (b) {$\mathbf{q}=1_m$};
				\node [below] at (b) {$\mathbf{\dot{x}} = \mathbf{1_n}$};
				\draw[-latex] (0,-1.65) arc(0:-225:0.5);
				\node [above] at (-0.6, -2.2) {$q\lambda$};
				\draw[-latex]  (-0.94, -1.2) -- (-2.08, -1.2);
				\node [below] at (-1.5, -1.2) {$\mu$};
				\draw[-latex]  (0.92, -0.8) -- (2.08, -0.8);
				\node [above] at (1.5, -0.9) {$\lambda_c$};
				\draw[-latex]  (0, -0.35) -- (0, 0.35);
				\node [left] at (0, 0) {$p\lambda$};
				
				\coordinate (c) at (0,3);
				\filldraw[fill=green!30!blue!50, line width=0.6pt] (c) ellipse[x radius=1, y radius=0.65]; 
				\node [above] at (c) {$\mathbf{q}=1_e$};
				\node [below] at (c) {$\mathbf{\dot{x}} = k_1\mathbf{x}$};
				\draw[-latex] (0,3.65) arc(0:225:0.5);
				\node [below] at (-0.6, 4.2) {$\lambda_c$};
				\draw[-latex]   (0.8, 2.6) --(2.2, 1.4);
				\node [right] at (1.5, 1.6) {$\lambda$};
				\draw[-latex]   (-1, 3) --(-2.5, 1.55);
				\node [left] at (-2.2, 1.8) {$\mu_c$};
				
				\coordinate (d) at (0,-3);
				\draw[line width=0.6pt] (d) ellipse[x radius=1, y radius=0.65]; 
				\node [above] at (d) {$\mathbf{q}=1_c$};
				\node [below] at (d) {$\mathbf{\dot{x}} = \mathbf{0_n}$};
				\draw[-latex] (0,-3.65) arc(0:-225:0.5);
				\node [right] at (0, -4) {$\lambda_c+q\lambda$};
				\draw[-latex]  (-1, -3)--(-2.6, -1.6);
				\node [left] at (-2.3, -2) {$\mu_c$};
				\draw[-latex]  (1, -3)--(2.6, -1.6);
				\node [left] at (1.6, -2.4) {$p\lambda$};
				
				\coordinate (e) at (3,1);
				\filldraw[fill=green!30!blue!80, line width=0.6pt] (e) ellipse[x radius=1, y radius=0.65]; 
				\node [above] at (e) {$\mathbf{q}=2_u$};
				\node [below] at (e) {$\mathbf{\dot{x}} = k_2\mathbf{x}$};
				\draw[-latex] (3,1.65) arc(180:-45:0.5);
				\node [above] at (4, 2) {$\lambda_c+\lambda$};
				\draw[-latex]  (2.08, 0.8) -- (0.92, 0.8);
				\node [below] at (1.5, 0.8) {$\mu_c$};
				\draw[-latex]  (2.6, 1.6)-- (1, 3) ;
				\node [right] at (1.1,3) {$\mu$};

				\coordinate (f) at (3,-1);
				\filldraw[fill=green!10!blue!10, line width=0.6pt] (f) ellipse[x radius=1, y radius=0.65]; 
				\node [above] at (f) {$\mathbf{q}=2_m$};
				\node [below] at (f) {$\mathbf{\dot{x}} = \mathbf{1_n}$};
				\draw[-latex] (3,-1.65) arc(-180:45:0.5);
				\node [below] at (4, -2.1) {$\lambda_c+q\lambda$};
				\draw[-latex]  (2.08, -1.2)--(0.94, -1.2);
				\node [below] at (1.5, -1.2) {$\mu_c$};
				\draw[-latex]  (3, -0.35) -- (3, 0.35);
				\node [right] at (3, 0) {$p\lambda$};
				\draw[-latex]  (2.2, -0.6) -- (0.6, 2.5);
				\node [left] at (0.9, 2) {$\mu$};
				
				\coordinate (g) at (-3,1);
				\filldraw[fill=green!30!blue!50, line width=0.6pt] (g) ellipse[x radius=1, y radius=0.65]; 
				\node [above] at (g) {$\mathbf{q}=0_e$};
				\node [below] at (g) {$\mathbf{\dot{x}} = k_1\mathbf{x}$};
				\draw[-latex]  (-2, 1) -- (-1, 1);
				\node [above] at (-1.5, 0.9) {$\lambda$};
				\draw[-latex]  (-2.2, 1.4) -- (-0.8, 2.6);
				\node [below] at (-1, 2.5) {$\lambda_c$};
				
				\coordinate (h) at (-3,-1);
				\draw[line width=0.6pt] (h) ellipse[x radius=1, y radius=0.65]; 
				\node [above] at (h) {$\mathbf{q}=0_c$};
				\node [below] at (h) {$\mathbf{\dot{x}} = \mathbf{0_n}$};
				\draw[-latex] (-3,-1.65) arc(0:-225:0.5);
				\node [above] at (-3.6, -2.2) {$q\lambda$};
				\draw[-latex]  (-2.08, -0.8) -- (-0.92, -0.8);
				\node [above] at (-1.5, -0.9) {$p\lambda$};
				\draw[-latex]  (-2.2, -1.4) -- (-0.8, -2.6);
				\node [above] at (-1, -2.5) {$\lambda_c$};
		\end{tikzpicture}}
		\caption{Graphical representation of SHSs: two M/M/1/1 queues over collision channel. The mappings $\phi_{\mathit{l}}$ is omitted for simplicity.}
		\label{SHS state2}
	\end{figure}
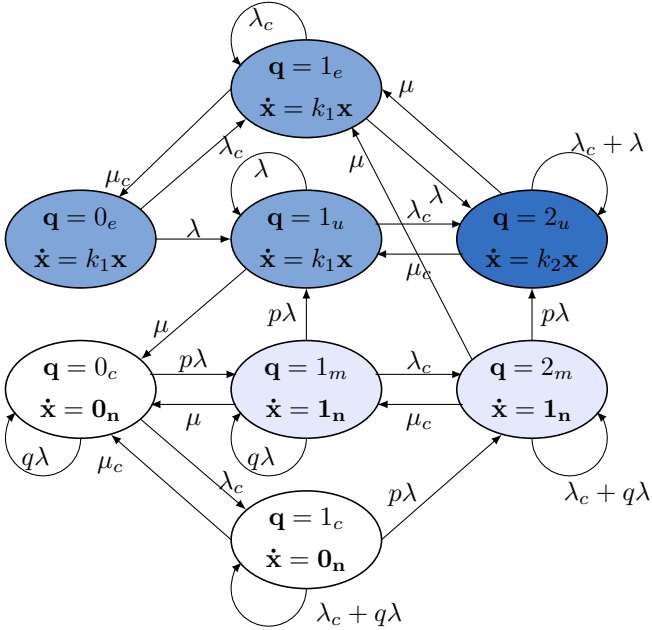
	\subsubsection*{Positive Recurrence}
	The properties of  positive recurrence and ergodicity in the collision channel case can be established by the same argument in the noisy channel case, hence, the proof here is omitted for simplicity.

	The SHSs states and corresponding transition set $\mathcal{L}_c$ are depicted in Fig. \ref{SHS state2}.  We now explain each transition $l\in\mathcal{L}_c$. 
	
	$l=0,6,9,13,16,20,24$: First, the assumption of the self-transition at state $0_c$ remains the same as the noisy channel case. In the remaining self-transitions, states $1_c$, $1_m$ and $2_m$ will stay unchanged when the transmission of updates is preempted either by a content-unchanged update  $X_t$ or the contender $X^c_t$. Moreover, states $1_u$, $1_e$ and $2_u$ will stay unchanged under any preemption.
	\begin{table}[htbp]
		\caption{State Transitions\label{table2}}
		\centering
		\renewcommand{\arraystretch}{0.6}
		\scalebox{0.6}{
			\begin{tabular}{cccccc}
				\toprule[1pt]
				$l$ & $\mathbf{q}_l\to \mathbf{q}_l^+$ & $\lambda^{(l)}$  & $\mathbf{x}\mathbf{A}_l$ & $\mathbf{A}_l$	& $\mathbf{v}_{q_l}\mathbf{A}_l$ \\
				\midrule
				$0$ & $0_c\to 0_c$ & $q\lambda$  & $\begin{bmatrix} 0 & 0 \end{bmatrix}$ & $\begin{bmatrix} 0 & 0 \\ 0 & 0 \end{bmatrix}$ &  $ \begin{bmatrix} 0 & 0 \end{bmatrix}$ \\
				
				$1$ & $0_c\to 1_c$ & $\lambda_c$  & $\begin{bmatrix} 0 & 0 \end{bmatrix}$ & $\begin{bmatrix} 0 & 0 \\ 0 & 0 \end{bmatrix}$ &  $ \begin{bmatrix} 0 & 0 \end{bmatrix}$ \\
				
				$2$ & $0_c\to 1_m$ & $p\lambda$  & $\begin{bmatrix} 0 & x_2 \end{bmatrix}$ & $\begin{bmatrix} 0 & 0 \\ 0 & 1 \end{bmatrix}$ &  $ \begin{bmatrix} 0 & v^2_{0_c} \end{bmatrix}$ \\			
				
				$3$ & $0_e\to 1_e$ & $\lambda_c$  & $\begin{bmatrix} x_2 & x_2 \end{bmatrix}$ & $\begin{bmatrix} 0 & 0 \\ 1 & 1 \end{bmatrix}$ &  $ \begin{bmatrix} v^2_{0_e} & v^2_{0_e} \end{bmatrix}$ \\
				
				$4$ & $0_e\to 1_u$ & $\lambda$  & $\begin{bmatrix} 0 & x_2 \end{bmatrix}$ & $\begin{bmatrix} 0 & 0 \\ 0 & 1 \end{bmatrix}$ &  $ \begin{bmatrix} 0 & v^2_{0_e} \end{bmatrix}$ \\
				
				$5$ & $1_c\to 0_c$ & $\mu_c$  & $\begin{bmatrix} 0 & 0 \end{bmatrix}$ & $\begin{bmatrix} 0 & 0 \\ 0 & 0 \end{bmatrix}$ &  $ \begin{bmatrix} 0 & 0 \end{bmatrix}$ \\					
				
				$6$ & $1_c\to 1_c$ & $\lambda_c+q\lambda$  & $\begin{bmatrix} 0 & 0 \end{bmatrix}$ & $\begin{bmatrix} 0 & 0 \\ 0 & 0 \end{bmatrix}$ &  $ \begin{bmatrix} 0 & 0 \end{bmatrix}$ \\		
				
				$7$ & $1_c\to 2_m$ & $p\lambda$  & $\begin{bmatrix} x_2 & x_2 \end{bmatrix}$ & $\begin{bmatrix} 0 & 0 \\ 1 & 1 \end{bmatrix}$ &  $ \begin{bmatrix} v^2_{1_c} & v^2_{1_c} \end{bmatrix}$ \\
				
				$8$ & $1_m\to 0_c$ & $\mu$  & $\begin{bmatrix} 0 & 0 \end{bmatrix}$ & $\begin{bmatrix} 0 & 0 \\ 0 & 0 \end{bmatrix}$ &  $ \begin{bmatrix} 0 & 0 \end{bmatrix}$ \\
				
				$9$ & $1_m\to 1_m$ & $q\lambda$  & $\begin{bmatrix} 0 & x_2 \end{bmatrix}$ & $\begin{bmatrix} 0 & 0 \\ 0 & 1 \end{bmatrix}$ &  $ \begin{bmatrix} 0 & v^2_{1_m}\end{bmatrix}$ \\
				
				$10$ & $1_m\to 2_m$ & $\lambda_c$  & $\begin{bmatrix} x_2 & x_2 \end{bmatrix}$ & $\begin{bmatrix} 0 & 0 \\ 1 & 1 \end{bmatrix}$ &  $ \begin{bmatrix} v^2_{1_m} & v^2_{1_m} \end{bmatrix}$ \\
				
				$11$ & $1_m\to 1_u$ & $p\lambda$  & $\begin{bmatrix} 0 & x_2 \end{bmatrix}$ & $\begin{bmatrix} 0 & 0 \\ 0 & 1 \end{bmatrix}$ &  $ \begin{bmatrix} 0 & v^2_{1_m} \end{bmatrix}$ \\
				
				$12$ & $1_u\to 0_c$ & $\mu$  & $\begin{bmatrix} 0 & 0 \end{bmatrix}$ & $\begin{bmatrix} 0 & 0 \\ 0 & 0 \end{bmatrix}$ &  $ \begin{bmatrix} 0 & 0 \end{bmatrix}$ \\			
				
				$13$ & $1_u\to 1_u$ & $\lambda$  & $\begin{bmatrix} 0 & x_2 \end{bmatrix}$ & $\begin{bmatrix} 0 & 0 \\ 0 & 1 \end{bmatrix}$ &  $ \begin{bmatrix} 0 & v^2_{1_u} \end{bmatrix}$ \\
				
				$14$ & $1_u\to 2_u$ & $\lambda_c$  & $\begin{bmatrix} x_2 & x_2 \end{bmatrix}$ & $\begin{bmatrix} 0 & 0 \\ 1 & 1 \end{bmatrix}$ &  $ \begin{bmatrix} v^2_{1_u} & v^2_{1_u} \end{bmatrix}$ \\
				
				$15$ & $1_e\to 0_e$ & $\mu_c$  & $\begin{bmatrix} x_2 & x_2 \end{bmatrix}$ & $\begin{bmatrix} 0 & 0 \\ 1 & 1 \end{bmatrix}$ &  $ \begin{bmatrix} v^2_{1_e} & v^2_{1_e} \end{bmatrix}$ \\
				
				$16$ & $1_e\to 1_e$ & $\lambda_c$  & $\begin{bmatrix} x_2 & x_2 \end{bmatrix}$ & $\begin{bmatrix} 0& 0 \\ 1 & 1 \end{bmatrix}$ &  $ \begin{bmatrix} v^2_{1_e} & v^2_{1_e} \end{bmatrix}$ \\
				
				$17$ & $1_e\to 2_u$ & $\lambda$  & $\begin{bmatrix} x_2 & x_2 \end{bmatrix}$ & $\begin{bmatrix} 0 & 0 \\ 1 & 1 \end{bmatrix}$ &  $ \begin{bmatrix} v^2_{1_e} & v^2_{1_e} \end{bmatrix}$ \\
				
				$18$ & $2_m\to 1_m$ & $\mu_c$  & $\begin{bmatrix} 0 & x_2 \end{bmatrix}$ & $\begin{bmatrix} 0 & 0 \\ 0 & 1 \end{bmatrix}$ &  $ \begin{bmatrix} 0 & v^2_{2_m} \end{bmatrix}$ \\
				
				$19$ & $2_m\to 1_e$ & $\mu$  & $\begin{bmatrix} x_2 & x_2 \end{bmatrix}$ & $\begin{bmatrix} 0 & 0 \\ 1 & 1 \end{bmatrix}$ &  $ \begin{bmatrix} v^2_{2_m} & v^2_{2_m} \end{bmatrix}$ \\
				
				$20$ & $2_m\to 2_m$ & $\lambda_c+q\lambda$  & $\begin{bmatrix} x_2 & x_2 \end{bmatrix}$ & $\begin{bmatrix} 0 & 0 \\ 1 & 1 \end{bmatrix}$ &  $ \begin{bmatrix} v^2_{2_m} & v^2_{2_m} \end{bmatrix}$ \\
				
				$21$ & $2_m\to 2_u$ & $p\lambda$  & $\begin{bmatrix} x_2 & x_2 \end{bmatrix}$ & $\begin{bmatrix} 0 & 0 \\ 1 & 1 \end{bmatrix}$ &  $ \begin{bmatrix} v^2_{2_m} & v^2_{2_m} \end{bmatrix}$ \\
				
				$22$ & $2_u\to 1_u$ & $\mu_c$  & $\begin{bmatrix} 0 & x_2 \end{bmatrix}$ & $\begin{bmatrix} 0 & 0 \\ 0 & 1 \end{bmatrix}$ &  $ \begin{bmatrix} 0 & v^2_{2_u} \end{bmatrix}$ \\
				
				$23$ & $2_u\to 1_e$ & $\mu$  & $\begin{bmatrix} x_2 & x_2 \end{bmatrix}$ & $\begin{bmatrix} 0 & 0 \\ 1 & 1 \end{bmatrix}$ &  $ \begin{bmatrix} v^2_{2_u} & v^2_{2_u} \end{bmatrix}$ \\
				
				$24$ & $2_u\to 2_u$ & $\lambda_c+\lambda$  & $\begin{bmatrix} x_2 & x_2 \end{bmatrix}$ & $\begin{bmatrix} 0 & 0 \\ 1 & 1 \end{bmatrix}$ &  $ \begin{bmatrix} v^2_{2_u} & v^2_{2_u} \end{bmatrix}$ \\
				\bottomrule[1pt]
			\end{tabular}
		}
	\end{table}
	\begin{figure*}
		\begin{subequations}
			\begin{footnotesize}
				\begin{equation}
					\begin{aligned}\label{collision distribution}
						&\pi^\prime_{0_c}=\frac{m\mu\mu_c(p\lambda+\mu_c)}{b(\lambda_c+\mu_c)},\ \pi^\prime_{1_c}=\frac{m\lambda_c\mu\mu_c}{b(\lambda_c+\mu_c)},
						\pi^\prime_{0_e}=\frac{p\lambda_c\mu\mu_c(m-q\lambda)}{b(\lambda_c+\mu_c)},\ \pi^\prime_{1_e}=\frac{p\lambda_c\mu(\lambda_c+\lambda)(m-q\lambda)}{b(\lambda_c+\mu_c)}\\
						&\pi^\prime_{1_m}=\frac{mp\lambda\mu\mu_c\big[(p\lambda+\mu+\mu_c)(p\lambda+\mu_c)+\lambda_c\mu_c\big]}{b(\lambda_c+\mu_c)(p\lambda+\mu)(m-q\lambda)},\
						\pi^\prime_{2_m}=\frac{mp\lambda_c\lambda\mu\mu_c(m-q\lambda)+mp^2\lambda_c\lambda^2\mu\mu_c}{b(\lambda_c+\mu_c)(p\lambda+\mu)(m-q\lambda)},\\
						&\pi^\prime_{1_u}=\frac{mp\lambda\mu_c\Big\{\big[p\lambda(p\lambda+\lambda_c+\mu_c)+\lambda_c\mu\big](m-q\lambda)+p\lambda_c\lambda\mu\Big\}}{b(\lambda_c+\mu_c)(p\lambda+\mu)(m-q\lambda)},\ \pi^\prime_{2_u}=1-\sum_{\mathbf{q}\in\mathbb{Q}_c/\{2_u\}}\pi_{\mathbf{q}}.
					\end{aligned}
				\end{equation}
			\end{footnotesize}
			\begin{footnotesize}
				\begin{equation}\label{quadratic average AoII2}
					\begin{aligned}
						x_c=\frac{1}{\Pi}&\Biggl\{c_2\lambda^2(\lambda_c+\mu+\mu_c-k_1)\big[p(m-2k_1)+\mu\big]+c_1p\lambda^2\big[(\lambda-k_1)(\lambda_c+\mu_c-k_2)+(\lambda_c+\mu_c-k_1)(\lambda_c+\mu+\mu_c-k_2)\big]\\
						&-c_2\lambda\biggl\{p(\lambda_c+\mu_c-k_1)\big[k_1(\lambda_c+\mu_c-k_1)-\mu(\lambda_c-2k_1)\big]-\mu^3-\mu(\lambda_c+\mu_c-k_1-k_2)\big[\mu(p+2)+\lambda_c+\mu_c-k_1\big]\biggr\}\\
						&-c_1p\lambda(\lambda_c+\mu_c-k_1)\big[\mu(\lambda_c-k_1)-k_1(\lambda_c+\mu_c-k_2)\big]\\
						&+c_2\mu(\lambda_c+\mu_c-k_1)\big[\mu^2+\mu(\lambda_c+\mu_c-k_1-k_2)-k_1(\mu_c-k_2)-k_2\lambda_c\big]
						\Biggr\}+c_1+c_2.
					\end{aligned}
				\end{equation}
			\end{footnotesize}
			\begin{footnotesize}
				\begin{equation}\label{determinant}
					\Pi=\det(\mathbf{D})(\lambda+\mu_c-k_1)(\lambda+\lambda_c-k_1)(\lambda_c+\mu-k_1)(\mu+\mu_c-k_2).
				\end{equation}
			\end{footnotesize}
			
			%	\begin{equation}
				%		\det(D)=\frac{\lambda^2\big[k_2(\lambda_c+\mu-k_1)-\mu_c(\mu-k_1)\big]-\lambda\biggl\{\mu^2(\mu_c-k_1)+\mu\big[k_1^2-k(\lambda_c+3\mu_c-2k_2)+(\mu_c-k_2)(\lambda_c+\mu_c)\big]+(\lambda_c+\mu_c-2k_1)\big[k_1(\mu_c-k_2)+k_2\lambda_c\big]\biggr\}+k_1(\lambda_c+\mu_c-k_1)\big[\mu(\lambda+\mu+\mu_c-k_1-k_2)-k_1(\mu_c-k_2)-k_2\lambda_c\big]}{(\lambda+\mu_c-k_1)(\lambda+\lambda_c-k_1)(\lambda_c+\mu-k_1)(\mu+\mu_c-k_1)}
				%	\end{equation}
		\end{subequations}
	\end{figure*}
	
	$l=1,2$: Since a newly-sampled update $X^c_t$ has no effect on the AoII of $X_t$ without collisions, the transition $l=1$ keeps $ x_1 = x_2 =0$. However, A newly-sampled update $X_t$ with changed content leads to mismatch, and the AoII starts to grow after the state is transferred into $1_m$ by $l=2$.
	
	$l=5,8,12$: The AoII  is reset to or remain zero after a successful transmission.
	
	$l=11, 21$: The update $X_t$ being transmitted is preempted by a new content-changed update. Hence, the mismatch  becomes severe. The states are transferred into $1_u$  and $2_u$ respectively. Note that the preemption has no effect on the channel collision in the transition $l=21$.

	$l=7, 10, 14, 17$:  When the channel is currently busy at transmitting the update $X^c_t$, collisions arise if a newly-sampled update $X_t$ with changed content arrives, i.e., $l=7$. Similarly,  for $l=17$, the channel collides for any newly-arrived update $X_t$ when $X^c_t$ is being transmitted. For $l=10, 14$, collisions arise  if any newly-sampled update of the contender arrives when the channel is currently busy at transmitting the update $X_t$.  The difference between $l=7, 10$ and $l=14, 17$ is in that the latter transfers the state into $2_u$, which suffers the fourth level age dissatisfaction due to a more severe content mismatch. Since neither update in transmission can be successfully decoded during the collision period, these four transitions all set $\mathbf{x}\mathbf{A}_l=\begin{bmatrix} x_2 & x_2 \end{bmatrix}$.
	
	$l=18, 19, 22, 23$: The collision ends if either $X_t$ ($l=18,22$) or $X^c_t$ ($l=18,23$) finishes transmission.   The difference between $l=18, 22$ and $l=19, 23$ is in that the latter transfers the state into $1_e$, where no update of  $X_t$ is being transmitted yet the mismatch of content still exists. This leads to   $\mathbf{x}\mathbf{A}_l=\begin{bmatrix} x_2 & x_2 \end{bmatrix}$ instead of  $\mathbf{x}\mathbf{A}_l=\begin{bmatrix} 0 & x_2 \end{bmatrix}$ in $l=18,22$.
	
	$l=3, 15$: The transitions $l=3, 15$ between states $0_e$ and $1_e$ only depend on the contender source, which cannot change the AoII of $X_t$. Moreover, the AoII keeps growing at the third level in these two error states because the content mismatch still exists.
	
	$l=4$: If the system were in the error state $0_e$, any newly-sampled update $X_t$ transfers the system state into $1_u$ and starts a transmission. 
	
	The state transitions above are summarized in Table \ref{table2}, along with the possible changes of AoII after mappings $\{\phi_{\mathit{l}}: l\in\mathcal{L}_c\}$.
	\subsection{Average AoII under Collision Channels}
	For collision case 1), the stationary distributions of the SHSs Markov chain in Fig. \ref{SHS state2} are calculated as \eqref{collision distribution}, where $m=\lambda_c+\lambda+\mu_c+\mu$ and $b=p(m\lambda+\lambda_c\mu)(m-q\lambda)+m\mu\mu_c$.\footnote{The explicit expression of $\pi_{2_u}$ cannot be easily factorized, which is too long to be included in \eqref{collision distribution}. Hence, we choose to present the implicit form.}
	
	Based on these, we obtain the average AoII of the collision channel case under a four-level hierarchy using Theorem \ref{theorem1}.
	\theoremstyle{plain}\newtheorem{result3}[corollarycounter]{Corollary}
	\begin{result3}\label{corollary3}
		For hierarchy scheme 3, the average AoII $x_c$ of two M/M/1/1 queues over the collision channel at the monitor is summarized as \eqref{quadratic average AoII2} given that the steady-state stability conditions satisfy, where the denominator is given in \eqref{determinant}. 
	\end{result3} 
	The derivation of Corollary \ref{corollary3} appears in Appendix \ref{app e}, where  $c_1$ and $c_2$ are given in \eqref{13} as the expected values of AoII at states $1_m$ and $2_m$, respectively. The associated matrix $\mathbf{D}$ is defined in \eqref{matrix3}, where the explicit expression of $\det(\mathbf{D})$ is given in Appendix \ref{app f} and is omitted here for simplicity.
	\subsection{Sufficient Conditions of Steady-state Stability} 
	For collision case 1),
	the stability conditions of  the general case \eqref{quadratic average AoII2} for any $\lambda,\lambda_c,\mu,\mu_c$ are hard to obtain since  $k_1$ and $k_2$ depend on the relationship between two pairs $(\lambda,\mu)$ and $(\lambda_c,\mu_c)$. Therefore, we now present some sufficient conditions for steady-state stability in the collision case 1) and give our finer results for a  symmetric case where the contender $X^c_t$ is evenly matched to $X_t$, i.e., $\lambda=\lambda_c$ and $\mu=\mu_c$.
	
	\theoremstyle{plain}\newtheorem{stability2}[theoremcounter]{Theorem}
	\begin{stability2}\label{theorem3}
		Assume that the following conditions are satisfied in the collision channel:
		
		i). The determinant of the associated matrix $\mathbf{D}$ is large than zero, i.e., $\det(\mathbf{D})>0$;
		
		ii). The growth constants $k_1$ and $k_2$ is limited, i.e., $0<k_1<\lambda$ and $0<k_1<k_2<\mu$.
		
		Then \eqref{quadratic average AoII2} exists for any $\lambda,\lambda_c,\mu,\mu_c$.
		
		Furthermore, if $\lambda=\lambda_c$ and $\mu=\mu_c$, the above conditions can be modified as:
		\begin{equation}\label{modified condition}
			\begin{aligned}
				&	0<k_1<\lambda<\mu,\\
				&	0<k_1<k_2<\min\{\mu, \frac{2k_1\mu^2+4k_1\lambda\mu-2k_1^2\mu-2\lambda\mu^2}{(3\lambda+\mu)k_1-k_1^2-\lambda\mu-2\lambda^2}\}.
			\end{aligned}
		\end{equation}
	\end{stability2}
	
	The proof of Theorem \ref{theorem3} appears in Appendix \ref{app f}. Note that \eqref{modified condition} helps to reveal the relationship between $k_1$ and $k_2$ similar to Theorem \ref{theorem2} with the knowledge of the rates of contender.
	
	\theoremstyle{remark}\newtheorem{remark4}[remarkcounter]{Remark}
	\begin{remark4} \label{remark4}
		We recognize that the range of $k_1$ and $k_2$ ensuring stability might be (much) larger than the intervals given in Theorem \ref{theorem3}, which are affected by  the contender with unknown parameters. However, the relationship between $k_1$ and $k_2$ given by condition i) is intractable unless we know the relationship of rates between the contender and the source of interest. Nevertheless, it is sufficient to adopt some appropriate constants given in the above conditions to achieve two main goals: a). The ability to distinguish different requirements of freshness at different system states; b). The ability to ensure the steady-state stability of real-time stochastic communication systems with limited consideration of the effects from environment. In practice, we prefer to choose $k_1$ and $k_2$ smaller than $\lambda$ which are independent of the rates of contender. 
	\end{remark4}
	
	\subsection{Discussions on Collision Case 2)} 
	We provide the analysis of collision case 2), where no updates can be successfully decoded after suffering collisions, in the supplementary material of this paper.
	\section{Numerical Results}\label{numerical}
	In this section,  we compare the classical results over the M/M/1/1 queue using traditional AoI and AoII metric. Meanwhile, numerical results of the average hierarchical AoII over the noisy and collision channels are given under different parameters, respectively. The effects of different channel conditions are also analyzed.

	\subsection{Comparisons between AoI and AoII: The M/M/1/1 queue}\label{Comparisons between AoI and AoII: The M/M/1/1 queue}
	To begin, we first need to introduce several results obtained in the M/M/1/1 queue using AoI, where the parameters $\lambda$ and $\mu$ are arrival and service rates of updates.

	1). \textit{The M/M/1/1 queue}: an update is accepted in the system only if the server is idle, furthermore, new arrivals are discarded when there is an update in service. \textit{The average AoI}  is given by \cite[Equ. (21)]{costa_age_2016}, i.e.,
	\begin{equation}\label{M/M/1/1}
		\Delta_{M/M/1/1}=\frac{1}{\lambda}+\frac{2}{\mu}-\frac{1}{\lambda+\mu}.
	\end{equation}
	
	2). \textit{The M/M/1/1 queue with preemption}: an update-in-service is preempted and discarded if there is a new arrival. \textit{The average AoI} is given by \cite[Equ. (8)]{najm_status_2018}\footnote{also see \cite[Equ. 18]{soysal_age_2021}.}, where the source only generates updates of one kind and the service time is set to be exponential, i.e.,
	\begin{equation}\label{M/M/1/1 preemption}
		\Delta^p_{M/M/1/1}=\frac{1}{\lambda}+\frac{1}{\mu}.
	\end{equation}

	3). \textit{The M/M/1/1 queue with abandonment}: an update-in-service is abandoned (i.e. discarded without completing service) at rate $\alpha$.  \textit{The average AoI} is given by \cite[Equ. (24)]{yates_age_2020b},
	\begin{equation}\label{M/M/1/1 abandonment}
		\Delta^\alpha_{M/M/1/1}=\frac{1}{\lambda}+\frac{1}{\mu}+\frac{\lambda}{(\mu+\alpha)(\lambda+\mu+\alpha)}+\frac{\alpha}{\lambda\mu}.
	\end{equation}
	Note that if we set the abandonment rate $\alpha=0$, then we have the traditional case in \eqref{M/M/1/1}.
	
	Since the channel of the above AoI results is assumed to be noise-free, we need to modify our AoII result in \eqref{simplified linear AoII} for a fair comparison by setting $p_e=0$, which results in:
	
	4). \textit{The average AoII  of the M/M/1/1 queue with preemption}:
	\begin{equation}\label{M/M/1/1 AoII}
		x^\ast_{a}=\frac{p\lambda}{(p\lambda+\mu)\mu}=\frac{1}{\mu}-\frac{1}{p\lambda+\mu}.
	\end{equation}
	Meanwhile, the AoS results over the noisy-free channel can obtained from \eqref{M/M/1/1 AoII} by letting $p=1$, that is,
	
	5). \textit{The average AoS  of the M/M/1/1 queue with preemption}:
	\begin{equation}\label{AoS}
		x^\ast_{a,AoS}=\frac{\lambda}{(\lambda+\mu)\mu}=\frac{1}{\mu}-\frac{1}{\lambda+\mu}.
	\end{equation}
	
	Based on these results, we first observe that
	\begin{equation}
		\Delta^p_{M/M/1/1}>\frac{2}{\mu},\quad 	x^\ast_{a,AoS}<\frac{1}{2\mu},
	\end{equation}
	under the assumption $\lambda<\mu$. Then, we have the  relationship  in the preemption $M/M/1/1$ queue among three different age metrics, that is,
	\begin{equation}\label{inequality1}
		\Delta^p_{M/M/1/1}> 4x^\ast_{a,AoS}> 4x^\ast_{a}.
	\end{equation} 
	The reason of the first inequality is that the AoI metric incorporates a redundant ageing process when the content of update is unchanged at the transmitter.  The reason of the second inequality is that the possible change of content ($p<1$) using AoII prolongs the occupancy time of state $0_c$ compared to the AoS case ($p=1$). Since the age at $0_c$ is always zero, the prolongation helps to further reduce the average.
	
	Interestingly, we also find that
	\begin{equation}
		\Delta_{M/M/1/1}=\Delta^p_{M/M/1/1}+x^\ast_{a,AoS}.
	\end{equation}
	Combined with \eqref{inequality1}, we have
	\begin{equation}\label{inequality2}
		\Delta_{M/M/1/1}>5x^\ast_{a,AoS}>5x^\ast_{a}.
	\end{equation}
	Equations \eqref{inequality1} and \eqref{inequality2}  indicates that the ceasing of age growth at the matched state sharply cuts down the average age in the M/M/1/1 queue systems, which in turn enables a large gain on the system performance of information freshness.
	
	Last but not least, we note that
	\begin{equation}
		x^\ast_{a,AoS}<\frac{1}{\mu},
	\end{equation}
	where the constant $\frac{1}{\mu}$ can be interpreted as the average age of a generate-at-will queue with Poisson service rate $\mu$. A generate-at-will queue is the queue that a new content-changed update arrives exactly when the current update finishes transmission. Hence, this inequality reveals that the occupancy time of state $0_c$ in \eqref{AoS} brings a reduction of $\frac{1}{\lambda+\mu}$ in the average. Moreover, under the same $p$, we can further conclude that the relationship of average age between \eqref{M/M/1/1 AoII} and the generate-at-will queue depends on $p$ and the utilization factor $\rho$.
	
	\begin{figure}[htbp] 
		\centering
		\includegraphics[width=0.48\textwidth]{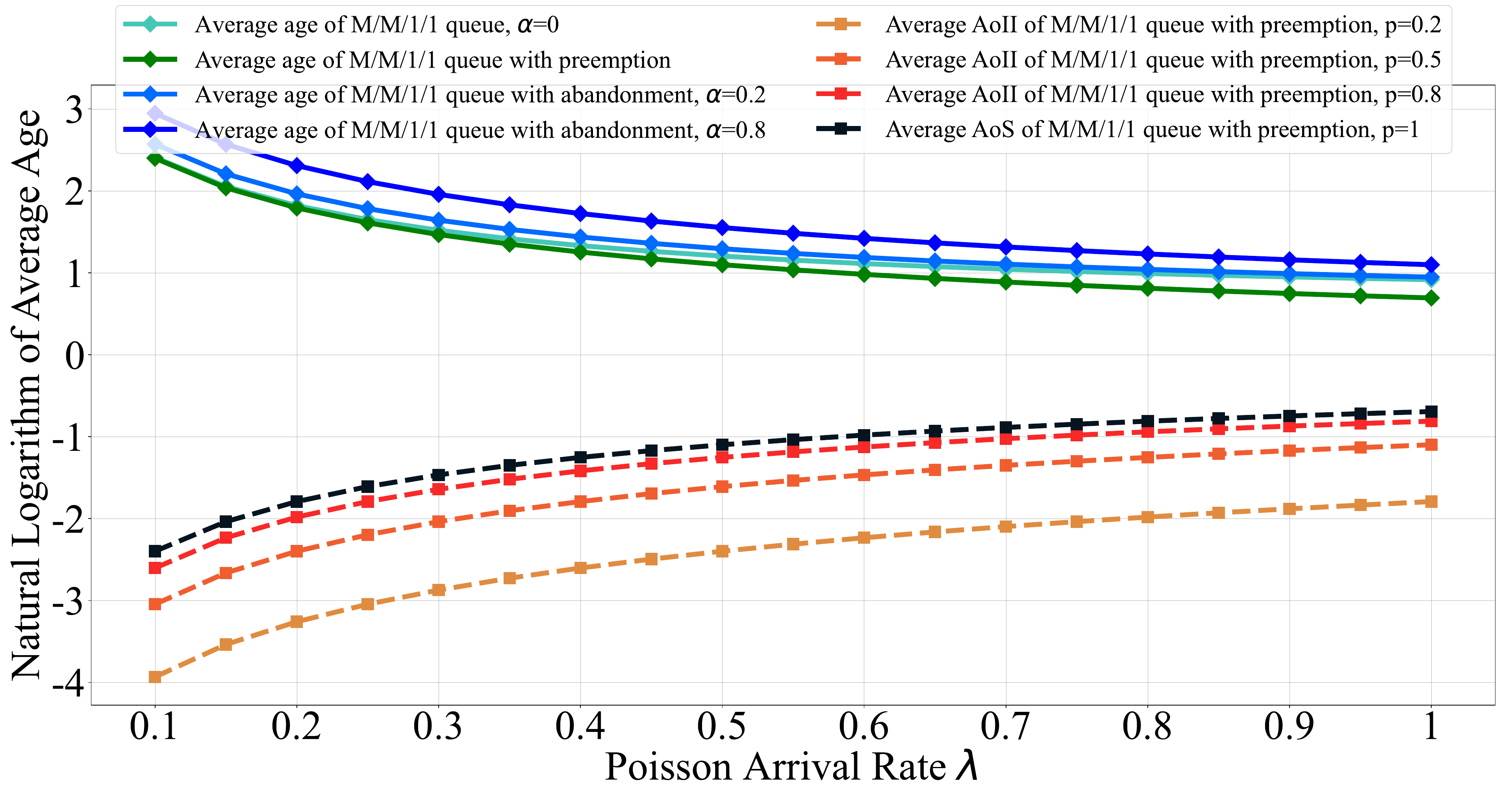}
		\caption{Illustration of the comparisons between the average AoI and AoII over the M/M/1/1 queue.}
		\label{fig1}
	\end{figure}
	
	We illustrate the above results numerically in Fig. \ref{fig1} by setting the service rate $\mu=1$, where we adopt the natural logarithm of the average. Note that a larger abandonment rate $\alpha$ leads to a larger average AoI, and a larger content-change probability $p$ gives a larger average AoII. 
	
	More interestingly, the average AoI is monotonically decreasing with the Poisson arrival rate $\lambda$ (equivalently, the utilization factor $\rho$) yet the average AoII is monotonically increasing with $\lambda$. The reason for this paradox lies in the inherent difference between two age metrics. To be specific, a larger arrival rate in the AoII-based M/M/1/1 queue shortens the occupancy time in $0_c$, which in turn gives a larger average system age due to the frequently changed content. On the contrary, in the AoI-based M/M/1/1 queue, a larger arrival rate shortens the occupancy time when no update is on transmission yet the AoI still grows, which ensures a smaller average system age.
		\begin{figure*}[htbp]
		\centering
		\subfloat[]{\includegraphics[width=0.3\textwidth]{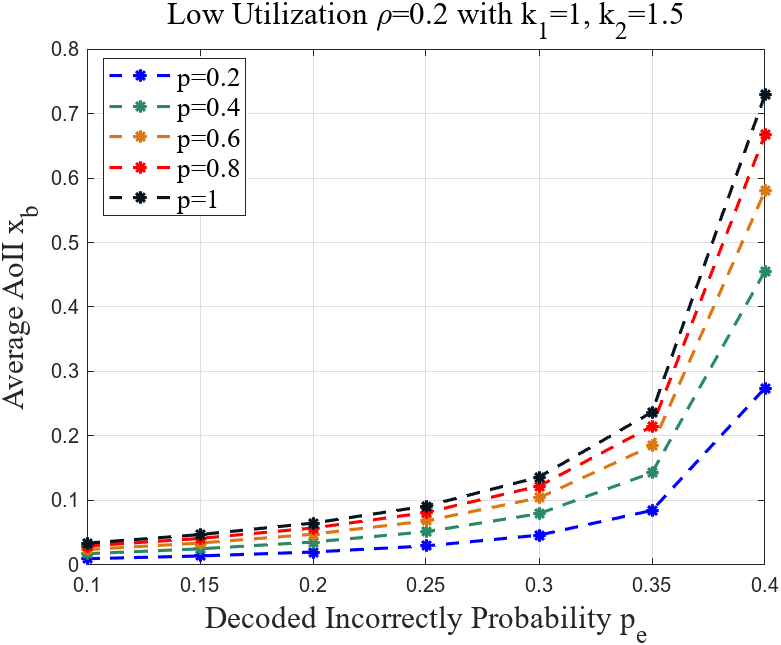}
			\label{B1}}
		\centering
		\subfloat[]{\includegraphics[width=0.3\textwidth]{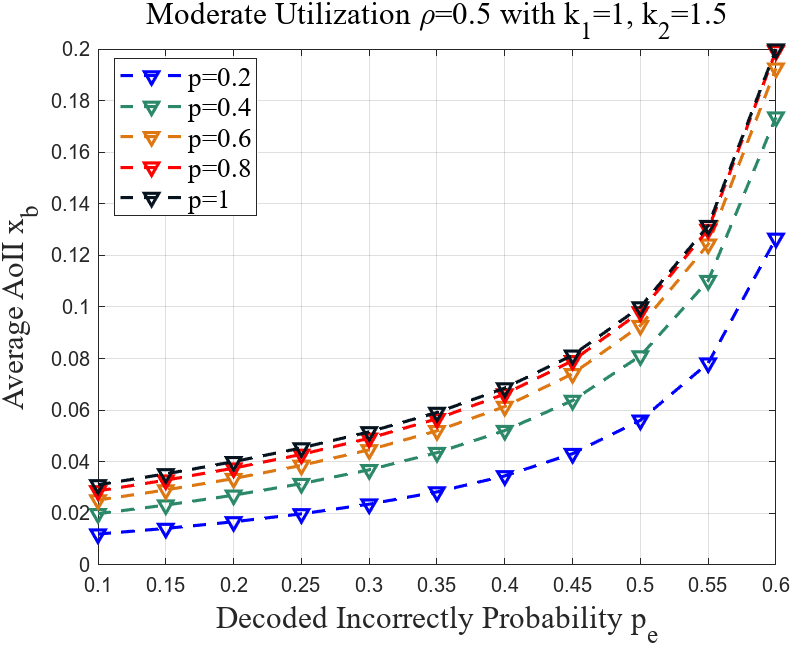}
			\label{B2}}
		\centering
		\subfloat[]{\includegraphics[width=0.3\textwidth]{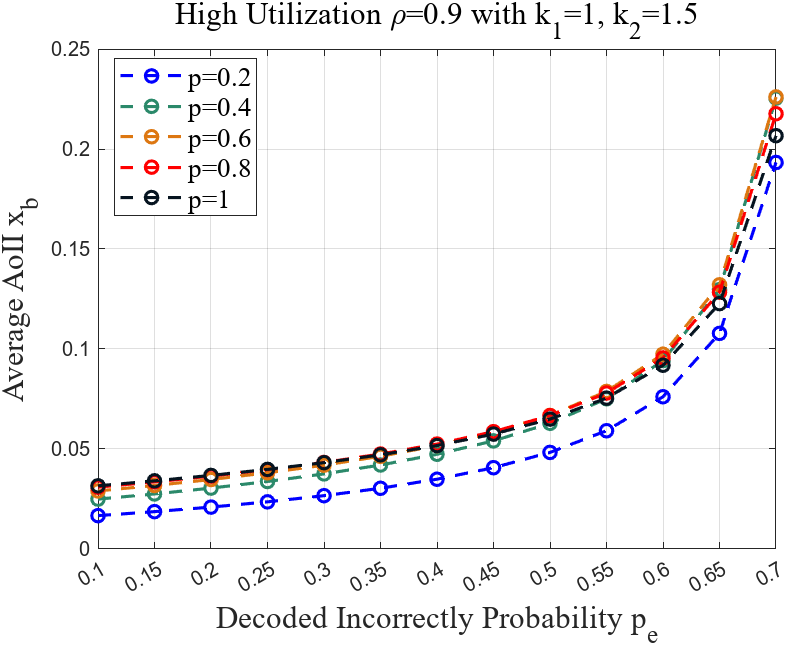}
			\label{B3}}
		\caption{The average AoII  over noisy channel with changing parameters $p$ and $p_e$, where the growth constants $k_1=1$ and $k_2=1.5$. The curves under different utilization factors $\rho=0.2, 0.5, 0.9$ are given in (a), (b), (c), respectively.}\label{fig2}
	\end{figure*}
	
	\subsection{The effects of noise in the M/M/1/1 queue}

	The curves of \eqref{quadratic average AoII} are depicted in Fig. \ref{fig2} with $k_1=1$ and $k_2=1.5$. Two main effects of channel noise on the average AoII can be concluded:
	
	1). The average AoII grows monotonically with $p_e$ regardless of utilization factors $\rho$ and content-changed probabilities $p$ in Fig. \ref{fig2}. The reason is obvious since the mismatch between ends cannot be timely eliminated due to large $p_e$.
	
	2). The range of $p_e$ is limited, e.g., $p_e<0.4$, $p_e<0.6$ and $p_e<0.7$ in Fig. \ref{B1}, \ref{B2} and \ref{B3}, respectively, due to the steady-state stability conditions given in Theorem \ref{theorem2}. Based on this, we conclude that a poor channel condition may lead to instability even the growth rates are relatively small. In SHSs, it means that any $p_e$ violating the stability conditions results in a finite escape time of the AoII with non-zero probability.
	
	Furthermore, when the channel error probability $p_e$ is large, e.g., $p_e>0.5$, the average AoII  does not always monotonically increase in $p$ under the high utilization $\rho=0.9$ as seen in Fig. \ref{B3}. One possible reason of this inconsistency is that the occupation time in state $0_e$/$1_u$ for $p=1$ is longer/shorter compared to that of $p=0.8$ or $p=0.6$  under the high $\rho$, which can be derived from \eqref{linear stationary distribution}. Since the growth constant $k_2$ being in state $1_u$ is larger than the growth constant $k_1$  being in state $0_e$, the average AoII for $p=1$ is lower instead.
	
	\subsection{The effects of collisions in two M/M/1/1 queues}
\begin{figure}[htbp] 
		\centering
		\includegraphics[width=0.38\textwidth]{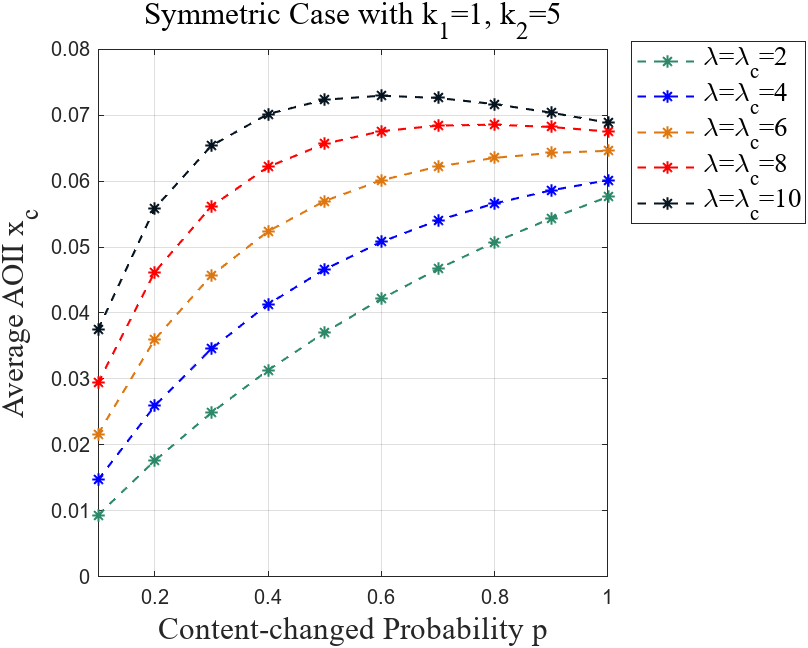}
		\caption{The average AoII over the collision channel of case 1), where  $k_1=1$ and $k_2=5$ are chosen to satisfy the stability conditions in Theorem \ref{theorem3}.}
		\label{B4}
	\end{figure}

	Considering the complexity of \eqref{quadratic average AoII2}, we only illustrate the symmetric case, i.e., $\lambda=\lambda_c$ and $\mu=\mu_c$, where the stability conditions in Theorem \ref{theorem3} are easy to obtain. The curves of different values of utilization factors (with $\mu=\mu_c=10$ fixed) are depicted in  Fig. \ref{B4} with varying content-changed probability $p$. The average AoII increases monotonically with larger utilization factors as expected since the higher arrival rates of updates  from both sources are more likely to cause channel collisions.\footnote{The reason why the average AoII does not increase monotonically in $p$ under large $\lambda$ is still unclear as of this writing. Nevertheless, we believe that a reasonable explanation  still lies in the stationary distributions of discrete state $\mathbf{q}$.	To find the exact answer, one could calculate the derivative of $\pi^\prime_{2_u}$ in \eqref{collision distribution} with respect to $p$ to see if a higher $\lambda$ will or not lead to a longer occupancy time in $\pi^\prime_{2_u}$. However, considering that the explicit expression of $\pi_{2_u}$ cannot be easily factorized, it is  too complex to obtain the closed-form expression of the derivative.}
	
	\section{Discussions and Conclusions}\label{Discussions}
	We first introduce state-dependent AoII processes under different hierarchy schemes. We provide a systematic way to derive the closed-form expressions of the average  AoII and obtain explicit results under two different channel models.  Moreover, we give the stability analysis based on the SHSs, which are essential for the existence of the average AoII. 
	
	As an alternative of  renewal-reward methods in obtaining closed-form expressions, the SHSs method introduced by \cite{yates_age_2019} is considered to be more powerful in that it can not only deal with age-dependent transition rates, which is first studied in \cite{maatouk_age-aware_2023}, also the state-dependent ageing processes that are first considered in our work. However, the limitations of SHSs are also apparent as indicated in the collision channel case, i.e., the computation complexity of Theorem \ref{theorem1}. To be clear, one can discuss two simple cases of generalization:
	
	i). Variations of multi-level hierarchical AoII processes with a finer state definitions;
	
	ii). General multi-access cases with $M\geqslant3$ sources under a larger finite state space $\mathbb{Q}_M$.
	
	We have discussed i) after Theorem \ref{theorem2}, also in Remark \ref{remark3}. For ii), the state definition for the AoII-based multi-access is more complex than the AoI-based multi-access model considered in \cite{yates_age_2020a}, which in fact makes this generalization intractable. In other words, although both generalizations follow the same analysis procedure in this work, the computation cost for each  is enormous due to the order of the matrix associated with the AoII SHSs.  To be specific, if we considered a general multi-access case with $M\geqslant3$ sources, we need to track the number $k$ of sources that have updates transmitting via the channel, which causes a $k$-way collision as noted in \cite{yates_age_2020a}. However, by adopting the hierarchical AoII, the simplest definition of discrete states in $\mathbb{Q}_M$ is $\mathbb{Q}_M=\mathbb{S}_1\cup\mathbb{S}_2\cup\mathbb{S}_3$, where
	\begin{equation}
		\begin{aligned}
			\mathbb{S}_1&=\{0_e, 0_c\},\ \mathbb{S}_2=\{M_m, M_u\},\\
			\mathbb{S}_3&=\{k_c,k_m,k_u,k_e: 1\leqslant k\leqslant M-1\}.
		\end{aligned}
	\end{equation}
	The meanings of states are defined similarly as in the two-nodes collision model according to their indexes. Hence, the number of states in $\mathbb{Q}_M$ is at least $4M$, which in turn gives a set of linear equations with the size of the associated matrix being $3M\times3M$.\footnote{the number of irrelevant states is $M$.}
	Although we could solve for the above generalization with the help of computer software, the closed-form expressions of average AoII are very complex, and no stability conditions can be easily obtained. A promising direction on the analysis of continuous AoII in such complex stochastic queuing systems would be the searches for appropriate bounds.

	{\appendices
		\section{Proof of Theorem \ref{theorem1}: The Third Part} \label{app a}
		Since it follows from \eqref{test function} and Lemma \ref{main} that the change of expected values of $\psi_{\overline{\mathbf{q}}}$ is controlled by an individual stochastic differential equation at each state $\overline{\mathbf{q}}\in\mathbb{Q}$, we only need to rederive the formulas of expected values $\mathbf{v}_{\overline{\mathbf{q}}}$ using the linear growth rate function. 	Without loss of generality, we leverage state $1_u$ of the noisy channel case to prove \eqref{expect quadratic}.
		
		First, we need to represent the extended generator at ${1_u}$ as
		\begin{equation}\label{f1}
			\begin{aligned}
				(L\psi)_{1_u}(\mathbf{q},\mathbf{x},t)
				=&k_2\mathbf{x}\delta_{1_u,\mathbf{q}}\\
				&+\sum_{l=1}^{m}\lambda_l\Big[\psi_{1_u}\big(\phi_{\mathit{l}}(\mathbf{q},\mathbf{x},t)\big)-\psi_{1_u}(\mathbf{q},\mathbf{x},t)\Big].
			\end{aligned}
		\end{equation}
		The first term in the right-hand side of \eqref{f1} follows from 
		\begin{equation}\label{jacobian}
			\begin{aligned}
				f^\prime_2(1_u)\frac{\partial\psi_{1_u}(\mathbf{q}, \mathbf{x}, t)}{\partial\mathbf{x}}
				&=f^\prime_2(1_u)\mathbf{I}_2\delta_{1_u,\mathbf{q}}\\
				&=k_2\mathbf{x}\delta_{1_u,\mathbf{q}},
			\end{aligned}
		\end{equation}
		where $\mathbf{I}_2$ is the $2\times2$ identity matrix. The second term in the right-hand side of \eqref{f1} can be divided into two parts according to \eqref{two transition sets}, that is,
		\begin{equation}
			\begin{aligned}
				&\sum_{l=1}^{m}\lambda_l\Big[\psi_{1_u}\big(\phi_{\mathit{l}}(\mathbf{q},\mathbf{x},t)\big)-\psi_{1_u}(\mathbf{q},\mathbf{x},t)\Big]\\=& \sum_{l\in\mathcal{L}^o_{1_u}}\lambda_l\delta_{1_u,\mathbf{q}}\Big[\psi_{1_u}\big(\mathbf{q}^+,\mathbf{x}^+\big)-\psi_{1_u}(\mathbf{q},\mathbf{x},t)\Big]\\
				&+\sum_{l\in\mathcal{L}^i_{1_u}}\lambda_l\delta_{\mathbf{q}_l,\mathbf{q}}\Big[\psi_{1_u}\big(\mathbf{q}^+,\mathbf{x}^+\big)-\psi_{1_u}(\mathbf{q},\mathbf{x},t)\Big]\\
				=& \sum_{l\in\mathcal{L}^o_{1_u}}\lambda_l\delta_{1_u,\mathbf{q}}\Big[\mathbf{x}\mathbf{A}_l\delta_{1_u,\mathbf{q}^+}-\mathbf{x}\delta_{1_u,\mathbf{q}}\Big]\\
				&+\sum_{l\in\mathcal{L}^i_{1_u}}\lambda_l\delta_{\mathbf{q}_l,\mathbf{q}}\Big[\mathbf{x}\mathbf{A}_l\delta_{1_u,\mathbf{q}^+}-\mathbf{x}\delta_{1_u,\mathbf{q}}\Big]\\
				=& -\mathbf{x}\delta_{1_u,\mathbf{q}}\sum_{l\in\mathcal{L}^o_{1_u}}\lambda_l+\sum_{l\in\mathcal{L}^i_{1_u}}\lambda_l\mathbf{x}\mathbf{A}_l\delta_{\mathbf{q}_l,\mathbf{q}},
			\end{aligned}
		\end{equation}
		where $\mathbf{q}_l$ is defined as the states  directly connected to $1_u$, and $\mathbf{q}$ and $\mathbf{q}^+$ are states right before and  after transitions at time $t$. Note that the second equality follows from \eqref{test function}, and the last equality follows because the self-transition has no contribution in both transition sets, which is the same for all states in our cases. Combined with \eqref{jacobian}, we have
		\begin{equation}
			(L\psi)_{1_u}
			=k_2\mathbf{x}\delta_{1_u,\mathbf{q}}-\mathbf{x}\delta_{1_u,\mathbf{q}}\sum_{l\in\mathcal{L}^o_{1_u}}\lambda_l+\sum_{l\in\mathcal{L}^i_{1_u}}\lambda_l\mathbf{x}\mathbf{A}_l\delta_{\mathbf{q}_l,\mathbf{q}}.
		\end{equation}
		Therefore, the expected value of the extended generator is
		\begin{equation}\label{f2}
			\mathrm{E}\big[(L\psi)_{1_u}\big]
			=\Big[k_2-\sum_{l\in\mathcal{L}^o_{1_u}}\lambda_l\Big]\mathbf{v}_{1_u}+\sum_{l\in\mathcal{L}^i_{1_u}}\lambda_l\mathbf{v}_{\mathbf{q}_l}\mathbf{A}_l.
		\end{equation}
		
		We can then substitute \eqref{f2} into Dynkin formula \eqref{diff dynkin}, i.e.,
		\begin{equation}
			\frac{\mathrm{d}\mathrm{E}[\psi_{1_u}(\mathbf{q}, \mathbf{x}, t)]}{\mathrm{d}t}= \Big[k_2-\sum_{l\in\mathcal{L}^o_{1_u}}\lambda_l\Big]\mathbf{v}_{1_u}+\sum_{l\in\mathcal{L}^i_{1_u}}\lambda_l\mathbf{v}_{\mathbf{q}_l}\mathbf{A}_l.
		\end{equation}
		
		Assuming that the stability condition is satisfied, the derivative shall equal to zero as $t\to\infty$. Hence,  we obtain
		\begin{equation}
			\mathbf{v}_{1_u}\Big[\sum_{l\in\mathcal{L}^o_{1_u}}\lambda_l-k_2\Big]=\sum_{l\in\mathcal{L}^i_{1_u}}\lambda_l\mathbf{v}_{\mathbf{q}_l}\mathbf{A}_l,
		\end{equation}
		and the proof is completed.
		
		\section{Proof of Positive Recurrence}\label{app a1}   
		Let $(\mathbf{q}_0, \mathbf{x}_0)$ denotes the initial condition at time $t_0$ and $\{t_i: i\geqslant0\}$ be the sequence of transition time. On each interval $[t_i, t_{i+1})$, the AoII at the monitor evolves as
		\begin{equation}\label{norm}
			x_2(s)=x_2(t_i)+\int_{t_i}^{s}f^\prime(\mathbf{q}(r), \mathbf{x}(r),r)\mathrm{d}r,\ \forall s \in [t_i, t_{i+1}),
		\end{equation}
		according to \eqref{flows}.  We first discuss Hierarchy Scheme 2.
		Setting $f^\prime=f^\prime_2$ and taking norms at both sides of \eqref{norm}, we have
		\begin{equation}\label{qu}
			||x_2(s)||\leqslant||x_2(t_i)||+\int_{t_i}^{s}\max\{k_2||x_2(r)||, c\}\mathrm{d}r,
		\end{equation}
		where $c=1$ is the given constant in our case\footnote{It is easy to verify that $c$ can be any positive constant.}. Since $k_2||x_2(r)||$ is continuous on each interval $[t_i, t_{i+1})$, one of the following cases must occur:
		
		I1).  $k_2||x_2(r)||<1$. Then we have
		\begin{equation}
			||x_2(s)||\leqslant||x_2(t_i)||+t_{i+1}-t_i. \forall s \in [t_i, t_{i+1}),
		\end{equation}
		
		I2).  $k_2||x_2(r)||>1$. Then we have
		\begin{equation}
			\begin{aligned}
				||x_2(s)||&\leqslant||x_2(t_i)||+\int_{t_i}^{s}k_2||x_2(r)||\mathrm{d}r\\
				&\leqslant \exp\Big[\int_{t_i}^{s}k_2\mathrm{d}r\Big] ||x_2(t_i)||, \forall s \in [t_i, t_{i+1}),\\
			\end{aligned}
		\end{equation}
		where the second inequality follows from the Bellman-Gronwall Lemma.
		
		I3).  $k_2||x_2(r_0)||=1$ for some $r_0\in[t_i, t_{i+1})$. Then we can leverage the Bellman-Gronwall Lemma on $r\in[r_0,t_{i+1})$ to obtain
		\begin{equation}
			\begin{aligned}
				||x_2(s)||&\leqslant \exp\Big[\int_{r_0}^{s}k_2\mathrm{d}r\Big] ||x_2(r_0)||, \ \forall s \in [r_0, t_{i+1}).
			\end{aligned}
		\end{equation}

		Combining I2) and I3),  for $\forall s \in [t_i, t_{i+1})$ we have
		\begin{equation}
			||x_2(s)||\leqslant \exp\Big[\int_{t_i}^{s}k_2\mathrm{d}r\Big] \max\{||x_2(t_i)||,  \frac{1}{k_2}\}.
		\end{equation}
		
		Considering the transition map $\phi_{\mathit{l}}$ defined in \eqref{mapping} at time $t_{i+1}$, we have
		\begin{equation}\label{10}
			\begin{aligned}
				||x_2(t_{i+1})||&=||\phi^x_l(\mathbf{\mathbf{q},\mathbf{x},t})||\\
				&\leqslant \exp\Big[\int_{t_i}^{t_{i+1}}k_2\mathrm{d}r\Big]\max\{||x_2(t_i)||,\frac{1}{k_2},0\},
			\end{aligned}
		\end{equation}
		where $\phi^x_l$ is the mapping of the continuous state. The inequality follows from the fact  that the AoII at the monitor either keeps growing or drops to zero at the transition time $t_{i+1}$.
		
		In the worst case, the second entry of initial condition $\mathbf{x}_0= (x_1(t_0), x_2(t_0))$ is larger than the constant $\frac{1}{k_2}$ and the first term in the maximum function of \eqref{10} always dominates as the AoII grows from $i=0$ to $i=k-1$, which yields
		\begin{equation}\label{11}
			||x_2(t_{k})||\leqslant \exp\Big[\int_{t_0}^{t_k}k_2\mathrm{d}r\Big]||x_2(t_0)||. 
		\end{equation}
		Given a positive probability of successful transmission $p_c$ in the noisy channel case, the discrete state $\mathbf{q}$ will eventually reach to $0_c$ in a finite time with probability one as $t\to\infty$. This condition can be easily verified by calculating the global balance equations for stationary distributions of $\mathbf{q}$. Denoting $t_k$ as the last transition time before $\mathbf{q}=0_c$, we obtain
		\begin{equation}
			\begin{aligned}
				||x_2(t)||&\leqslant \exp\Big[\int_{t_k}^{t}k_2\mathrm{d}r\Big] \max\{||x_2(t_k)||,  \frac{1}{k_2}\}\\
				&\leqslant \exp\Big[\int_{t_0}^{t}k_2\mathrm{d}r\Big] ||x_2(t_0)||, \forall t \in [t_k, t_{k+1}),\\
				&\leqslant \exp\Big[k_2(t_{k+1}-t_0)\Big] ||x_2(t_0)||,\\
				&< \infty,
			\end{aligned}
		\end{equation}
		where the last inequality follows from $t_{k+1}<\infty$. Therefore, $||x_2(t)||$ is always bounded and $x_2(t_{k+1})=0$. Denote $\mathbb{V}=(0_c,0)$ and  $(\mathbf{q}_0, \mathbf{x}_0)\in\mathbb{V}^c$. We can conclude that the AoII SHSs using $f^\prime_2$ under noisy channel is positive recurrent with respect to $\mathbb{V}=(0_c,0)$, the mean recurrence time of which is equal to that from $\mathbf{q}_0$ to $0_c$. The proof for the AoII SHSs using $f^\prime_1$ can be simply obtained by replacing $\max\{k_2||x_2(r)||, c\}$ in \eqref{qu} to the constant $m_2$ in $f^\prime_1$ hence omitted.

		\section{Derivations of Corollary \ref{corollary1}} \label{app b}
		Since we are only interested in the average AoII at the monitor, it is sufficient to track the expected value of the second entry of $\mathbf{v}_{\overline{\mathbf{q}}}$, namely, $v_{\overline{\mathbf{q}}}^2$ at each state $\overline{\mathbf{q}}$. Therefore, it turns out that \eqref{expect linear} becomes a set of linear equations,
		\begin{equation}
			\begin{aligned}
				\lambda v^2_{0_e}&=m_1 \pi_{0_e} + p_e\mu v^2_{1_m} + p_e\mu v^2_{1_u},\\
				(p\lambda+\mu)v^2_{1_m}&=\pi_{1_m} +p\lambda v^2_{0_c},\\
				\mu v^2_{1_u}&=m_2\pi_{1_u} +\lambda v^2_{0_e} +p\lambda v^2_{1_m},
			\end{aligned}
		\end{equation} 
		where $v^2_{0_c}$ equals to zero since the extended generator $(L\phi)_{0_c}$ is always zero. We reformulate the above equations into the matrix form $\mathbf{B}\mathbf{v}^\top=\mathbf{b}^\top_1$, i.e.,
		\begin{equation}\label{matrix1}
			\begin{bmatrix}
				1& -p_e\rho^{-1} &-p_e\rho^{-1}\\
				0& 1 & 0\\
				-\rho & -p\rho & 1
			\end{bmatrix}
			\begin{bmatrix}
				v^2_{0_e}\\	v^2_{1_m}\\ v^2_{1_u}
			\end{bmatrix}=
			\begin{bmatrix}
				\frac{m_1\pi_{0_e}}{\lambda}\\ \frac{\pi_{1_m}}{p\lambda +\mu} \\ \frac{m_2\pi_{1_u}}{\mu}
			\end{bmatrix},
		\end{equation}
		where $\mathbf{b}^\top_1$ is the constant column vector at the right-hand side of \eqref{matrix1}.  Hence, the solutions of the above linear equations are easy to obtain, that is,
		\begin{equation}
			\begin{aligned}
				v^2_{0_e}&=\frac{m_1pp_e}{a\lambda}+p_e\rho^{-1}(v^2_{1_m}+v^2_{1_u}),\\
				v^2_{1_m}&=\frac{pp_c}{a\lambda(p+\rho^{-1})^2},\\
				v^2_{1_u}&=\frac{1}{p_c\mu}\Big[\frac{m_2(p^2\rho+pp_e)}{a(p+\rho^{-1})}+\frac{m_1pp_e}{a}+\frac{pp_c(p_e\rho^{-1}+p)}{a(p+\rho^{-1})^2}\Big],
			\end{aligned}
		\end{equation} 
	}
	Substituting the above results into \eqref{average AoII}, we have \eqref{linear average AoII}. 
	
	\section{Derivations of Corollary \ref{corollary2}} \label{app c}
	By leveraging \eqref{expect linear} and \eqref{expect quadratic} at each state, we obtain the set of linear equations, i.e.,
	\begin{equation}
		\begin{aligned}
			(\lambda-k_1) v^2_{0_e}&= p_e\mu v^2_{1_m} + p_e\mu v^2_{1_u},\\
			(p\lambda+\mu)v^2_{1_m}&=\pi_{1_m} +p\lambda v^2_{0_c},\\
			(\mu-k_2) v^2_{1_u}&= \lambda v^2_{0_e} +p\lambda v^2_{1_m},
		\end{aligned}
	\end{equation}
	where $v^2_{0_c}$ equals to zero. We again reformulate the above equations into the matrix form $\mathbf{C}\mathbf{v}^\top=\mathbf{b}^\top_2$, i.e.,
	\begin{equation}\label{matrix2}
		\begin{bmatrix}
			1& \frac{p_e\mu}{k_1-\lambda} &\frac{p_e\mu}{k_1-\lambda} \\
			0& 1 & 0\\
			\frac{\lambda}{k_2-\mu} & \frac{p\lambda}{k_2-\mu} & 1
		\end{bmatrix}
		\begin{bmatrix}
			v^2_{0_e}\\	v^2_{1_m}\\ v^2_{1_u}
		\end{bmatrix}=
		\begin{bmatrix}
			0\\ \frac{\pi_{1_m}}{p\lambda +\mu} \\ 0
		\end{bmatrix}.
	\end{equation}
	
	Assuming that the stability conditions are satisfied where the inverse $\mathbf{C}^{-1}$ exists, the average AoII $x_b$ can be calculated by the summation of solutions of \eqref{matrix2}, that is,
	\begin{equation}\label{f3}
		x_b=\mathbf{1}_3 \mathbf{C}^{-1} \mathbf{b}^\top_2,
	\end{equation}
	where $\mathbf{1}_3=[1, 1, 1]$ and $\mathbf{b}^\top_2$ is the constant column vector at the right-hand side of \eqref{matrix2}.

	\section{Steady-state Stability Conditions of Hierarchy Scheme 2 over Noisy Channel} \label{app d}
	\subsubsection{Proof of Necessity}
	We first prove the necessity of these two conditions. Given that the joint state $(\mathbf{q}, \mathbf{x})$ is ergodic, the hierarchical AoII $x_2(t)$ converges to a limit $x_2$ that can be obtained by \eqref{f3}.  Therefore,  we are able to establish the necessary conditions from two aspects: 1) the invertibility of matrix $\mathbf{C}$; 2) the positivity of $v^2_{\overline{\mathbf{q}}}$ at each state $\overline{\mathbf{q}}$.
	
	Following 1) and 2), we need to calculate the related quantities from \eqref{matrix2}. The determinant of $\mathbf{C}$ is obtained by 
	\begin{equation}
		\det(\mathbf{C})= \sum_{j=1}^{3}c_{2j}M_{2j}=1-\frac{p_e\lambda\mu}{(k_1-\lambda)(k_2-\mu)}\neq0,
	\end{equation}
	where $\{c_{2j}:j=1,2,3\}$ is the elements in the second row of $\mathbf{C}$ and $M_{2j}$ is the cofator of the element  $c_{2j}$. Furthermore, the expected value $v^2_{\overline{\mathbf{q}}}$ of AoII can be calculated from \eqref{matrix2}, 
	\begin{equation}
		\begin{bmatrix}
			v^2_{0_e}\\	v^2_{1_m}\\ v^2_{1_u}
		\end{bmatrix}=\mathbf{C}^{-1} \mathbf{b}^\top_2=\frac{\pi_{1_m}}{p\lambda +\mu}
		\begin{bmatrix}
			(\mathbf{C}^{-1})_{12}\\	(\mathbf{C}^{-1})_{22}\\ 	(\mathbf{C}^{-1})_{32}
		\end{bmatrix},
	\end{equation}
	where the second column of $\mathbf{C}^{-1}$ can be obtained using the general formula for inverse matrix in linear algebra, i.e.,
	\begin{equation}
		\begin{aligned}
			(\mathbf{C}^{-1})_{12}&=\frac{M_{21}}{\det(\mathbf{C})}=\frac{\frac{p_e\mu}{\lambda-k_1}+\frac{pp_e\lambda\mu}{(k_1-\lambda)(k_2-\mu)}}{\det(\mathbf{C})},\\
			(\mathbf{C}^{-1})_{22}&=\frac{M_{22}}{\det(\mathbf{C})}=1,\\
			(\mathbf{C}^{-1})_{32}&=\frac{M_{23}}{\det(\mathbf{C})}=\frac{\frac{p\lambda}{\mu-k_2}+\frac{p_e\lambda\mu}{(k_1-\lambda)(k_2-\mu)}}{\det(\mathbf{C})}.
		\end{aligned}
	\end{equation} 
	As a result, we only need to discuss the relationships between two parameter pairs, namely, $(\lambda, k_1)$ and $(\mu, k_2)$, that ensure the invertibility as well as the positivity of $(\mathbf{C}^{-1})_{12}$ and $(\mathbf{C}^{-1})_{32}$, which are divided into two cases.
	
	\textit{- Case I:} $\det(\mathbf{C})>0$ and $M_{21}>0$ and $M_{23}>0$. In this case, we shall discuss the following four situations.
	
	I1). If $0<k_1<\lambda$, $0<k_2<\mu$, then we have
	\begin{subequations}
		\begin{equation}\label{0}
			(k_1-\lambda)(k_2-\mu)>p_e\lambda\mu,
		\end{equation}
		\begin{equation}\label{1}
			\mu-k_2+p\lambda>0,
		\end{equation}
		\begin{equation}\label{2}
			p(\lambda-k_1)+p_e\mu>0,
		\end{equation}
	\end{subequations}
	where \eqref{0}, \eqref{1} and \eqref{2} follow from $\det(\mathbf{C})>0$, $M_{12}>0$ and $M_{23}>0$, respectively. The latter two equations can be further simplified as
	\begin{subequations}
		\begin{equation}\label{3}
			k_2<\mu+p\lambda,
		\end{equation}
		\begin{equation}\label{4}
			k_1<\lambda+\frac{p_e\mu}{p},
		\end{equation}
	\end{subequations}
	both of which \textit{satisfy} the assumptions in I1).
	
	I2). If $k_1>\lambda$, $k_2>\mu$, then we again have \eqref{0}, \eqref{1} and \eqref{2}. However, the requirement in \eqref{0} cannot be met since
	\begin{equation}
		\max_{k_1,k_2}	(k_1-\lambda)(k_2-\mu) <\big[(\lambda+\frac{p_e\mu}{p})-\lambda\big]\big[(p\lambda+\mu)-\mu\big]=p_e\lambda\mu,
	\end{equation}
	where the first inequality follows from \eqref{3} and \eqref{4} that violates the condition $\det(\mathbf{C})>0$. 
	
	I3). If $0<k_1<\lambda$, $k_2>\mu$, then the condition $M_{23}>0$ cannot be satisfied since it yields
	\begin{equation}
		p(\lambda-k_1)+p_e\mu<0,
	\end{equation}
	which requires $k_1>\lambda+\frac{p_e\mu}{p}$.
	
	I4). If $k_1>\lambda$, $0<k_2<\mu$, then the condition $M_{21}>0$ cannot be satisfied since it yields
	\begin{equation}
		\mu-k_2+p\lambda<0,
	\end{equation}
	which requires $k_2>\mu+p\lambda$.
	
	Next, we consider the other case.
	
	\textit{- Case II:} $\det(\mathbf{C})<0$ and $M_{21}<0$ and $M_{23}<0$. We also need to discuss the same four situations as case I.
	
	II1). If $0<k_1<\lambda$, $0<k_2<\mu$, then we have
	\begin{subequations}
		\begin{equation}\label{5}
			(k_1-\lambda)(k_2-\mu)<p_e\lambda\mu,
		\end{equation}
		\begin{equation}\label{6}
			\mu-k_2+p\lambda<0,
		\end{equation}
		\begin{equation}\label{7}
			p(\lambda-k_1)+p_e\mu<0,
		\end{equation}
	\end{subequations}
	where \eqref{5}, \eqref{6} and \eqref{7} follow from $\det(\mathbf{C})<0$, $M_{12}<0$ and $M_{23}<0$, respectively. The latter two equations can be further simplified as
	\begin{subequations}
		\begin{equation}\label{8}
			k_2>\mu+p\lambda,
		\end{equation}
		\begin{equation}\label{9}
			k_1>\lambda+\frac{p_e\mu}{p},
		\end{equation}
	\end{subequations}
	both of which \textit{violate} the assumptions in II1).
	
	II2). If $k_1>\lambda$, $k_2>\mu$, we again have \eqref{5}, \eqref{6} and \eqref{7}. However, the requirement in \eqref{5} cannot be met since
	\begin{equation}
		\min_{k_1,k_2}	(k_1-\lambda)(k_2-\mu) >\big[(\lambda+\frac{p_e\mu}{p})-\lambda\big]\big[(p\lambda+\mu)-\mu\big]=p_e\lambda\mu,
	\end{equation}
	where the first inequality follows from \eqref{8} and \eqref{9} that violates the condition $\det(\mathbf{C})<0$. 
	
	II3). If $0<k_1<\lambda$, $k_2>\mu$, then the condition $\det(\mathbf{C})<0$ cannot be satisfied since
	\begin{equation}
		\det(\mathbf{C})=1-\frac{p_e\lambda\mu}{(k_1-\lambda)(k_2-\mu)}>0,
	\end{equation}
	under our assumptions in II3) due to the negativity of $(k_1-\lambda)(k_2-\mu)$.

	II4). If $k_1>\lambda$, $0<k_2<\mu$, then the condition $\det(\mathbf{C})<0$ cannot be satisfied for the same reason as II3).
	
	To summarize, we have proved the necessity of the steady-state stability conditions consisting of  $0<k_1<\lambda$, $0<k_1<k_2<\mu$ and \eqref{0} as given in Theorem \ref{theorem2}.
	\subsubsection{Proof of Sufficiency} Following Appendix \ref{app a1}.
	\section{Derivations of Corollary \ref{corollary3}} \label{app e}
	By leveraging \eqref{expect linear} and \eqref{expect quadratic} at each state $\mathbf{q}\in\mathbb{Q}_c$, we obtain the set of linear equations, i.e.,
	\begin{equation}\label{12}
		\begin{aligned}
			(p\lambda+\lambda_c+\mu)v^2_{1_m}&=\pi^\prime_{1_m}+\mu_c v^2_{2_m},\\
			(p\lambda+\mu+\mu_c)v^2_{2_m}&=\pi^\prime_{2_m}+\lambda_c v^2_{1_m},\\
			(\lambda+\lambda_c-k_1)v^2_{0_e}&=\mu_c v^2_{1_e},\\
			(\lambda+\mu_c-k_1)v^2_{1_e}&=\mu v^2_{2_m}+\lambda_c v^2_{0_e}+\mu v^2_{2_u},\\
			(\lambda_c+\mu-k_1)v^2_{1_u}&=p\lambda v^2_{1_m}+\lambda v^2_{0_e}+\mu_c v^2_{2_u},\\
			(\mu+\mu_c-k_2)v^2_{2_u}&=p\lambda v^2_{2_m}+\lambda v^2_{1_e}+\lambda_c v^2_{1_u},
		\end{aligned}
	\end{equation}
	where $v^2_{0_c}=v^2_{1_c}=0$ is omitted above since $0_c$ and $1_c$ are irrelevant states. Note that the expected AoII at states $1_m$ and $2_m$ grows like the traditional AoI, which are only related to each other. These two equations can be solved first
	\begin{equation}\label{13}
		\begin{aligned}
			v^2_{1_m}=\frac{(p\lambda+\mu+\mu_c)\pi_{1_m}^\prime+\mu_c\pi_{2_m}^\prime}{(p\lambda+\lambda_c+\mu)(p\lambda+\mu+\mu_c)-\lambda_c\mu_c}=c_1,\\
			v^2_{2_m}=\frac{(p\lambda+\lambda_c+\mu)\pi_{2_m}^\prime+\lambda_c\pi_{1_m}^\prime}{(p\lambda+\lambda_c+\mu)(p\lambda+\mu+\mu_c)-\lambda_c\mu_c}=c_2,
		\end{aligned}
	\end{equation}
	and denote as constant $c_1$ and $c_2$, respectively. Therefore, \eqref{12} can be reduced to a set of $4\times4$ linear equations with the matrix form $\mathbf{D}\mathbf{v}^\top=\mathbf{b}^\top_3$, i.e.,
	\begin{equation}\label{matrix3}
		\begin{aligned}
			\begin{bmatrix}
				1& \frac{\mu_c}{k_1-\lambda-\lambda_c} &0 & 0\\
				\frac{\lambda_c}{k_1-\lambda-\mu_c}& 1 & 0 & \frac{\mu}{k_1-\lambda-\mu_c}\\
				\frac{\lambda}{k_1-\lambda_c-\mu} & 0 &1 &\frac{\mu_c}{k_1-\lambda_c-\mu} \\
				0 & \frac{\lambda}{k_2-\mu-\mu_c} & \frac{\lambda_c}{k_2-\mu-\mu_c} & 1
			\end{bmatrix}
			\begin{bmatrix}
				v^2_{0_e}\\ v^2_{1_e} \\	v^2_{1_u}\\ v^2_{2_u}
			\end{bmatrix}\\
			=
			\begin{bmatrix}
				0\\ \frac{\mu c_2 }{\lambda+\mu_c-k_1} \\ \frac{p\lambda c_1}{\lambda_c+\mu-k_1}\\ \frac{p\lambda c_2}{\mu+\mu_c-k_2}
			\end{bmatrix}.
		\end{aligned}
	\end{equation}
	
	Assuming that the stability conditions are satisfied where the inverse $\mathbf{D}^{-1}$ exists, the average AoII $x_c$ can be calculated by the summation of solutions of \eqref{13} and \eqref{matrix3}, that is,
	\begin{equation}\label{14}
		x_c=c_1+c_2+\mathbf{1}_4 \mathbf{D}^{-1} \mathbf{b}^\top_3,
	\end{equation}
	where $\mathbf{1}_4=[1, 1, 1,1]$ and $\mathbf{b}^\top_3$ is the constant column vector at the right-hand side of \eqref{matrix3}.

	\section{Sufficient Stability Conditions of Hierarchy Scheme 3 over the Collision Channel} \label{app f}
	The determinant of $\mathbf{D}$ is obtained in \eqref{determinant2} using the definition. The expected values of AoII at states $1_m,2_m$ are given in \eqref{13} as $c_1$ and $c_2$, both of which are larger than zero. Then we need to calculate the remaining expected values of AoII at states $0_e, 1_e, 1_u, 2_u$, which are summarized in \eqref{expected AoII2}.
	\begin{figure*}
		\begin{scriptsize}
			\begin{equation}\label{determinant2}
				\begin{aligned}
					\det(\mathbf{D})=&\frac{\Pi}{(\lambda+\mu_c-k_1)(\lambda+\lambda_c-k_1)(\lambda_c+\mu-k_1)(\mu+\mu_c-k_2)},\\
					\Pi=\biggl\{&\lambda\mu^2\mu_c+(\lambda\mu\mu_c-k_2\lambda\mu-k_2\lambda_c)(\lambda+\lambda_c+\mu_c)+(k_2-\mu-\mu_c)k_1^3\
					+\big[\mu(m-k_2)+(\mu+\mu_c)(\lambda_c+\mu_c-k_2)+2\lambda(\mu_c-k_2)-2k_2\lambda_c\big]k^2_1\\
					&-\big[(\lambda_c+\mu+\mu_c-k_2)(\lambda\mu+\lambda\mu_c+\lambda_c\mu+\mu\mu_c)+\lambda(\lambda+\mu)(\mu_c-k_2)-3k_2\lambda\lambda_c\big]k_1
					\biggr\}.
				\end{aligned}
			\end{equation}
		\end{scriptsize}
	\end{figure*}
	\begin{figure*}
		\begin{scriptsize}
			\begin{equation}\label{expected AoII2}
				\begin{aligned}
					v^2_{0_e}&=\frac{1}{\Pi}\biggl\{c_2\mu\mu_c A+c_1p\lambda\lambda_c\mu\mu_c\biggr\},\
					v^2_{1_e}=\frac{1}{\Pi}\biggl\{c_2\mu(\lambda+\lambda_c-k_1)A+c_1p\lambda\lambda_c\mu(\lambda+\lambda_c-k_1)\biggr\},\\
					v^2_{1_u}&=\frac{1}{\Pi}\biggl\{c_2\lambda\mu\mu_c(m-k_1-k_2)+(c_2p\lambda\mu_c+\frac{c_1p\lambda\lambda_c\mu_c}{\lambda_c+\mu-k_1})B+\frac{c_1p\lambda\Pi}{\lambda_c+\mu-k_1}\biggl\},\\
					v^2_{2_u}&=\frac{1}{\Pi}\biggl\{c_2\lambda\mu\big[\lambda_c(m-2k_1)+(\lambda-k_1)(\mu-k_1)\big]+p\lambda(\lambda-k_1)(\lambda+\lambda_c+\mu_c-k_1)\big[c_1\lambda_c+c_2(\lambda_c+\mu-k_1)\big]\biggr\},\\
					A&=(k_2-p\lambda-\mu-\mu_c)k_1-(\lambda_c+\mu)k_2+(\lambda_c+\mu+\mu_c)\mu+(\lambda_c+\mu)p\lambda,\
					B=k_1^2-(2\lambda+\lambda_c+\mu_c)k_1+m\lambda.
				\end{aligned}
			\end{equation}
		\end{scriptsize}
	\end{figure*}
	\begin{figure*}
		\begin{scriptsize}
			\begin{equation}\label{prime}
				\begin{aligned}
					C=2\mu^2(\lambda-k_1)+2k^2_1\mu-(4\lambda-k_2)k_1\mu-(2\lambda-k_1)(\lambda-k_1)k_2-k_2\lambda\mu,\ \Pi^\prime=(\lambda+\mu-k_1)C.
				\end{aligned}
			\end{equation}
		\end{scriptsize}
	\end{figure*}
	With the above quantities prepared, we are now ready to verify our sufficient conditions proposed in Theorem \ref{theorem3} by checking  the positivity of \eqref{expected AoII2} and the invertibility of $\mathbf{D}$. Given that $0<k_1<\lambda<\mu$ and $0<k_2<\mu$, we only need to verify the positivity of $A$ and $B$ defined in \eqref{expected AoII2} since the rest parts are easily checked to be positive. Consequently, we have the following discussions.
	
	I). Since $A$ is a monotonically decreasing linear function of $k_1$, we have
	\begin{equation}
		\begin{aligned}
			\min_{k_1} A &> (k_2-p\lambda-\mu-\mu_c)\lambda-(\lambda_c+\mu)k_2\\
			&+(\lambda_c+\mu+\mu_c)\mu+(\lambda_c+\mu)p\lambda \\
			&=(\lambda-\lambda_c-\mu)k_2 + (\lambda_c+\mu+\mu_c)\mu\\
			&+(\lambda_c+\mu)p\lambda-(p\lambda+\mu+\mu_c)\lambda\\
			&=A^\prime
		\end{aligned}
	\end{equation}
	where $A^\prime$ is again a monotonically decreasing linear function of $k_2$. Given $k_2<\mu$, we have
	\begin{equation}
		\begin{aligned}
			A^\prime&>(\lambda-\lambda_c-\mu)\mu + (\lambda_c+\mu+\mu_c)\mu\\
			&+(\lambda_c+\mu)p\lambda-(p\lambda+\mu+\mu_c)\lambda\\
			&=\mu\mu_c+(\lambda_c+\mu)p\lambda-(p\lambda+\mu_c)\lambda\\
			&=(\mu-\lambda)(p\lambda+\mu_c)+p\lambda\lambda_c\\
			&>0.
		\end{aligned}
	\end{equation}
	
	Hence, we can obtain
	\begin{equation}
		A>\min_{k_1} A>A^\prime>0,
	\end{equation}
	which guarantees the positivity of constant $A$.
	
	II). Constant $B$ is a parabola of $k_1$, whose direction of opening is upward. The focus of  this parabola is at point $(\lambda+\frac{\lambda_c+\mu_c}{2}, m\lambda-(\frac{\lambda_c+\mu_c}{2})^2)$. However, since the curve of $B$ is limited by the value of $k_1$, we can obtain the minimal of $B$ in the interval $k_1\in(0,\lambda)$, i.e.,
	\begin{equation}
		\begin{aligned}
			\min_{k_1} B>\lambda^2-(2\lambda+\lambda_c+\mu_c)\lambda+m\lambda
			=\lambda\mu
			>0,
		\end{aligned}
	\end{equation}
	where the equality follows from $m=\lambda+\lambda_c+\mu+\mu_c$. Hence, we obtain the positivity of constant $B$.
	
	Since we have the positivity of \eqref{expected AoII2}, combined with $\Pi$ in \eqref{determinant2} being positive, we obtain the sufficient stability conditions of the general case in Theorem \ref{theorem3}.
	
	Now, we can have a finer result of condition i) in Theorem \ref{theorem3} for the symmetric case where $\lambda=\lambda_c$ and $\mu=\mu_c$. To that end, we first simplify $\Pi$ as $\Pi^\prime$ in \eqref{prime}. Since we have
	\begin{equation}
		\begin{aligned}
			2\lambda-k_1>0,\
			2\mu-k_2>0,\
			\lambda+\mu-k_1>0,
		\end{aligned}
	\end{equation}
	using condition ii), we only need to guarantee the constant $C$ to be positive in order to satisfy condition i) under the symmetric case. As a result, we have
	\begin{equation}
		\begin{aligned}
			C=&Dk_2+2k^2_1\mu +2\lambda\mu^2-4k_1\lambda\mu-2k_1\mu^2,\\
			D=&-k_1^2+(3\lambda+\mu)k_1-\lambda\mu-2\lambda^2
		\end{aligned}
	\end{equation}
	
	Again, we have a parabola $D$ of  $k_1$  whose direction of opening is downward. Since the curve of this parabola is limited by $k_1\in(0,\lambda)$, the maximal of $D$ is less than the value when $k_1=\lambda$, i.e.,
	\begin{equation}
		\max_{k_1\in(0,\lambda)} D< D_{k_1=\lambda}=0.
	\end{equation}
	
	To ensure $\det(\mathbf{D})>0$, we need $C>0$, which results in
	\begin{equation}
		k_2<\frac{2k_1\mu^2+4k_1\lambda\mu-2k_1^2\mu-2\lambda\mu^2}{(3\lambda+\mu)k_1-k_1^2-\lambda\mu-2\lambda^2}. 
	\end{equation}
	Combined with the above results, we have completed the proof of Theorem \ref{theorem3}.

	\bibliography{Citation}
\end{document}